*Type of the Paper (Article)*

# Joint System Modeling Approach for Fault Simulation of Starter/Generator and Gas Generator in All-Electric APU

**Haotian Mao [1,*] and Yingqing Guo [2,*]**

[1] School of Power and Energy, Northwestern Polytechnical University, Xi'an 710072, China; alexmao@mail.nwpu.edu.cn
[2] School of Power and Energy, Northwestern Polytechnical University, Xi'an 710072, China; yqguo@nwpu.edu.cn
* Author to whom correspondence should be addressed.

**Abstract:** This paper presents a joint system modeling approach for fault simulation of all-electric auxiliary power unit (APU), integrating starter/generator turn-to-turn short circuit (TTSC) faults with gas generator gas-path faults. To address challenges in electromechanical coupling, simulation precision and computational efficiency balance, we propose a multi-rate continuous-discrete hybrid simulation architecture. This architecture treats the starter/generator as a continuous system with variable step size in Simulink, while modeling the gas generator as a discrete system with fixed step size in a dynamic-link library (DLL) environment. For the starter/generator fault modeling, a multi-loop approach is deployed to accurately simulate TTSC faults. For the gas generator, we develop an improved GasTurb-DLL modeling method (IGDM) that enhances uncertainty modeling, state-space representation, and tool chain compatibility. Finally, the proposed methodology above was implemented in a case study based on the APS5000 all-electric APU structure and parameters. Model validation was conducted by comparing simulation results—covering steady-state, transients, healthy, and fault conditions—with reference data from third-party software and literature. The close agreement confirms both the model's accuracy and the effectiveness of our modeling methodology. This work establishes a modeling foundation for investigating the opportunities and challenges in fault detection and isolation (FDI) brought by the all electrification of the APU, including joint fault estimation and diagnosis, coupled electromechanical fault characteristics.

**Keywords:** All-electric; auxiliary power unit (APU); more-electric; multi-rate simulation; turn-to-turn short circuit (TTSC); inter-turn faults; gas-path faults; starter/generator; gas generator; joint system modeling; fault simulation.

## 1. Introduction

The concept of aircraft electrification has gained significant momentum in recent years as the aviation industry seeks to reduce emissions, enhance efficiency, and improve reliability. In this context, the more electric aircraft (MEA) has emerged as a promising interim solution toward the ultimate goal of all-electric aircraft (AEA) [1-3]. As a key component in this electrification trend, the auxiliary power unit (APU) has undergone a substantial transformation, evolving from a traditional APU that provides electrical, pneumatic, and hydraulic power to an all-electric APU that exclusively supplies electrical power [3-8].

A representative example of this evolution is the Hamilton Sundstrand APS5000 all-electric APU, as shown in Figure 1 [9]. This all-electric APU deployed in the Boeing 787 Dreamliner, represent a significant advancement in aircraft power systems [1,3-5,10]. The all-electric APU eliminate the load compressor and hydraulic pumps found in traditional APU, replacing them with high-power starter/generators. This structural simplification results in weight reduction, improved fuel efficiency, and enhanced operational flexibility [4-6]. The all-electric APU primarily consists of two major components: a gas generator that operates as a turboshaft engine and a high-power starter/generator that converts mechanical power to electrical power. The operational mechanism involves the gas generator transmitting power to the starter/generator via a drive shaft and gearbox, enabling electricity generation to power various aircraft systems [4,7,11-13].

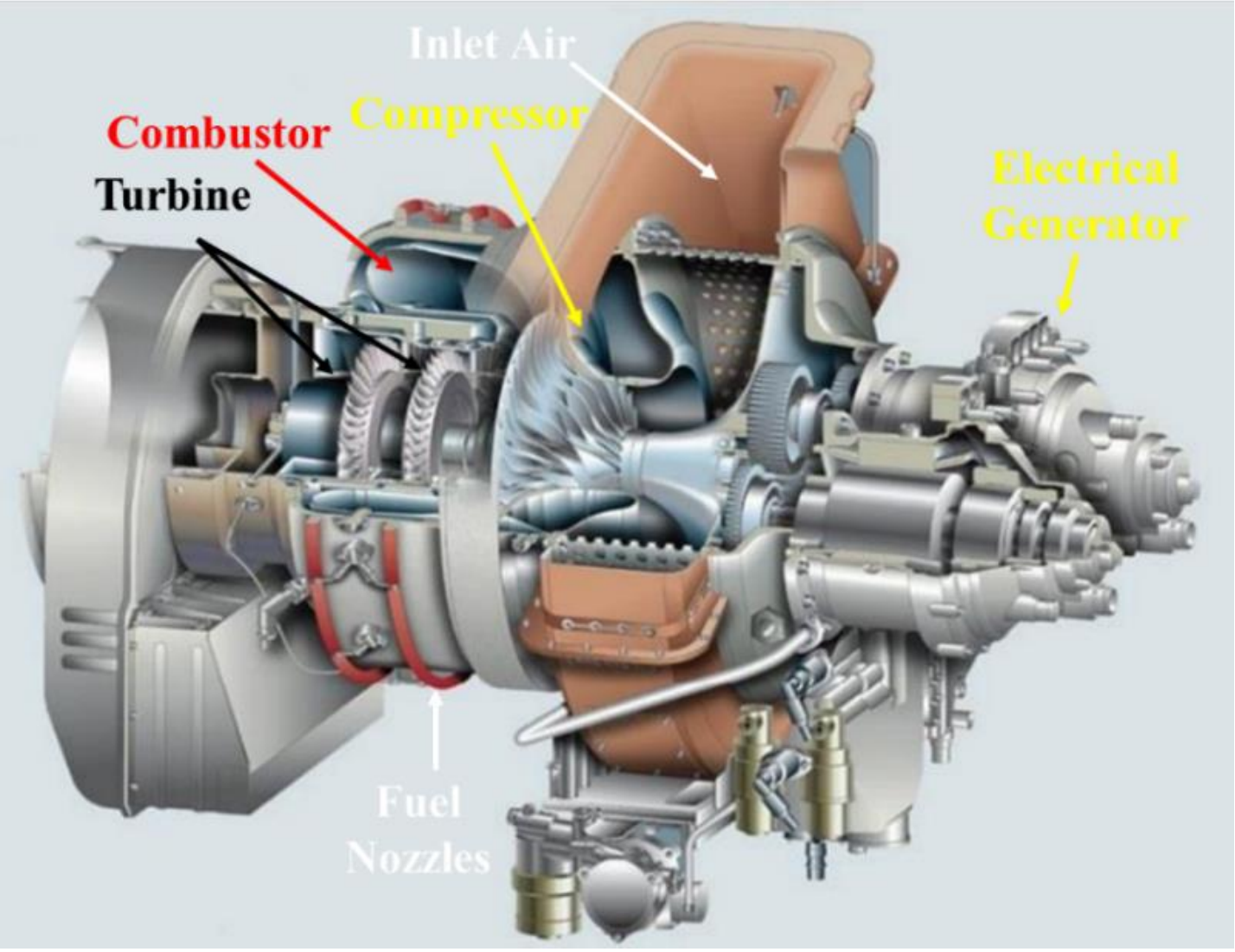

**Figure 1.** All-electric APU structure (Hamilton Sundstrand APS5000 [9])

The reliability and performance of all-electric APU are critical for flight safety and operational efficiency. The APU is essential for starting the main engines and providing power during ground operations, while also serving as a backup power source during flight [6,14,15]. Therefore, accurate fault detection, isolation, and prediction in all-electric APU are vital for ensuring flight safety and implementing condition-based maintenance strategies to reduce maintenance costs [4-6].

Among the various faults that can affect all-electric APU, stator turn-to-turn short circuit (TTSC) faults in the starter/generator and gas-path faults in the gas generator are particularly common, critical, and significant [16-21]. Stator TTSC faults can lead to numerous serious starter/generator malfunctions that are difficult to detect in their early stages, while gas-path faults directly impact gas generator performance and safety [16,22-26]. Accurate fault modeling is essential for understanding fault mechanisms, developing diagnostic methodologies, and validating detection algorithms, which has led to extensive research in this area [27-30].

For starter/generator stator TTSC faults, several modeling approaches have been proposed in the literature. The authors in [31] developed a hybrid model combining dq0 modeling with winding function approach (WFA) to improve modeling accuracy for TTSC faults. The authors in [32] created an extended dynamic model in qd0 stationary reference-frame with a vectorial fault factor to include interturn faults in any phase-windings. The authors in [33] employed a flux-linkage-based approach by extracting flux linkage maps from finite-element analysis to model synchronous motor with stator turn faults. The authors in [34] proposed a multi-loop model for fractional pole-path ratio synchronous generators, calculating mutual inductances between stator coils with arbitrary pitch and modeling core localized saturation effects by modifying the air gap function of fault coils.

Similarly, for gas-path faults in aviation gas turbine, numerous modeling techniques have been developed. The authors in [35] developed an integrated diagnostic model for gas path faults in turboshaft engines and employed back propagation (BP) and radial basis function (RBF) neural networks specifically for diagnosis. The authors in [36] develops a component-level gas path fault modeling approach for turboshaft engines by modifying health parameters to simulate specific component degradations, and then proposes two extreme learning machine-based transfer learning methods to improve fault diagnosis performance. The authors in [37] created an aircraft engine model capable of simulating gas path faults in MATLAB/Simulink environment by developing a GasTurb-DLL modeling method (GDM).

While these studies have made significant contributions to component-level fault modeling, there remains a gap in joint fault modeling that captures the interactions between starter/generator and gas generator faults in all-electric APU. The authors in [7] developed a component-level model of an all-electric APU for cooperative control purposes, but their work did not address fault conditions.

This gap is significant considering the strong coupling relationship that exists between these components in an all-electric APU. Compared to traditional APU, the all-electric APU reduces the loads on the gas generator from three to just one—the starter/generator—as shown in Figure 2, where the dashed lines represent components eliminated in the all-electric APU compared to traditional designs. This configuration creates a strong coupling relationship between the starter/generator and gas generator in all-electric APU, which presents both opportunities and challenges for fault analysis.

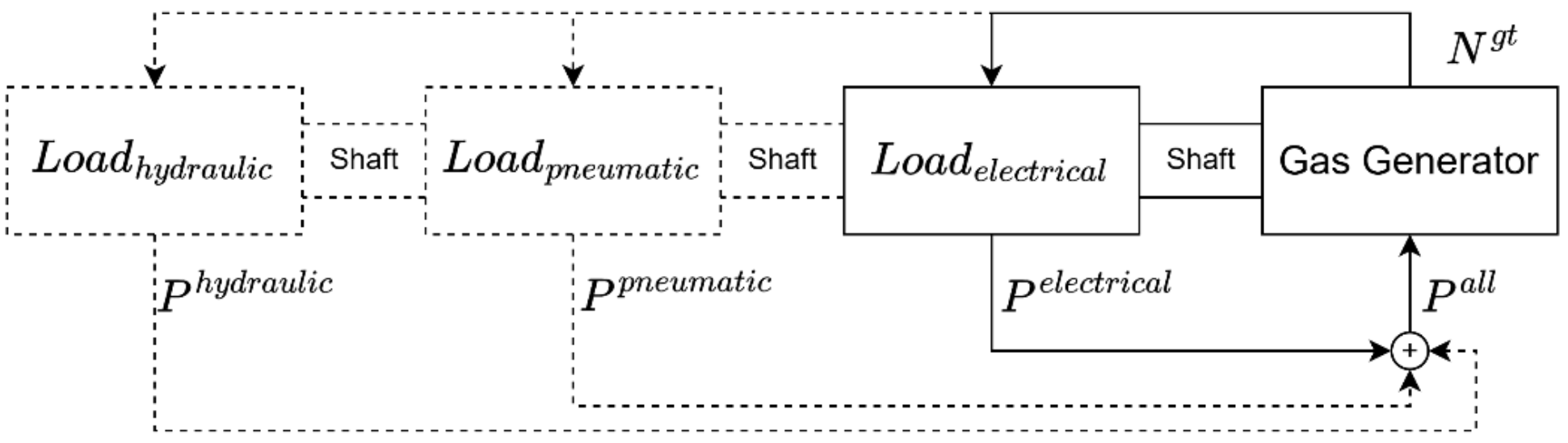

**Figure 2.** Comparison between the traditional APU and an all-electric APU

On one hand, the strong coupling relationship between the starter/generator and the gas generator, cause their faults to have significant impacts on each other and rendering starter/generator faults more critical [7, 17]. This presents a new challenge for our FDI approach. On the other hand, the strong coupling relationship also offers opportunities to leverage this relationship to enhance our FDI methodology and performance. By utilizing information from one system, one can improve the estimates of the other system.

In our previous study [38], we provided a detailed explanation of the above content. We also demonstrated that the electrification of the APU opens up new possibilities for utilizing shaft power estimates from the starter/generator to improve gas generator fault detection and isolation (FDI) performance. We established that in all-electric APU, where the shaft power output of gas generator is exclusively directed to the starter/generator rather than distributed among electrical, pneumatic, and hydraulic loads, accurate estimates of the gas generator's shaft power can be obtained through electrical measurements from the starter/generator. These estimates can significantly enhance the accuracy of gas generator fault estimation and improve overall FDI performance, particularly when using more accurate models or when estimating a greater number of health parameters.

To further validate these theoretical findings and investigate the potential of joint fault diagnosis in all-electric APU, a high-fidelity, nonlinear joint fault simulation model is required. This model will also provide support for future in-depth exploration of the new opportunities and challenges that all-electric APU bring to fault estimation and FDI.

Based on these motivations and research gaps, this paper proposes a novel joint system modeling approach for fault simulation of all-electric APU, integrating starter/generator TTSC faults with gas generator gas-path faults. The remainder of this paper is organized as follows: Section 2 introduces the research framework, including research motivations, problems, methodologies, innovations, and outcomes. Section 3 presents the joint modeling approach, proposing a multi-rate continuous-discrete hybrid simulation architecture that balances simulation precision with computational efficiency while accurately capturing the complex interactions between components. Section 4 demonstrates the starter/generator modeling, deploying a multi-loop method to establish a model capable of simulating both normal operation and TTSC faults. Section 5 showcases the gas generator modeling, developing an improved GasTurb-DLL modeling method (IGDM) that enhances uncertainty modeling, state-space representation, and tool chain compatibility. Section 6 provides validation results comparing our model with reference data from third-party software and literature. Finally, Section 7 concludes the paper and discusses future research directions.

## 2. Research Framework

Figure 3 illustrates the core content of this paper, highlighting the main technical innovations and the key problems they address.

As shown in Figure 3, the complexity of all-electric APU fault modeling and its electromechanical coupling characteristics arise three challenging technical problems: (1) balancing electromechanical coupling modeling with precision and computational load; (2) modeling the aviation starter/generator turn-to-turn short circuit faults; and (3) modeling gas generator gas path faults.

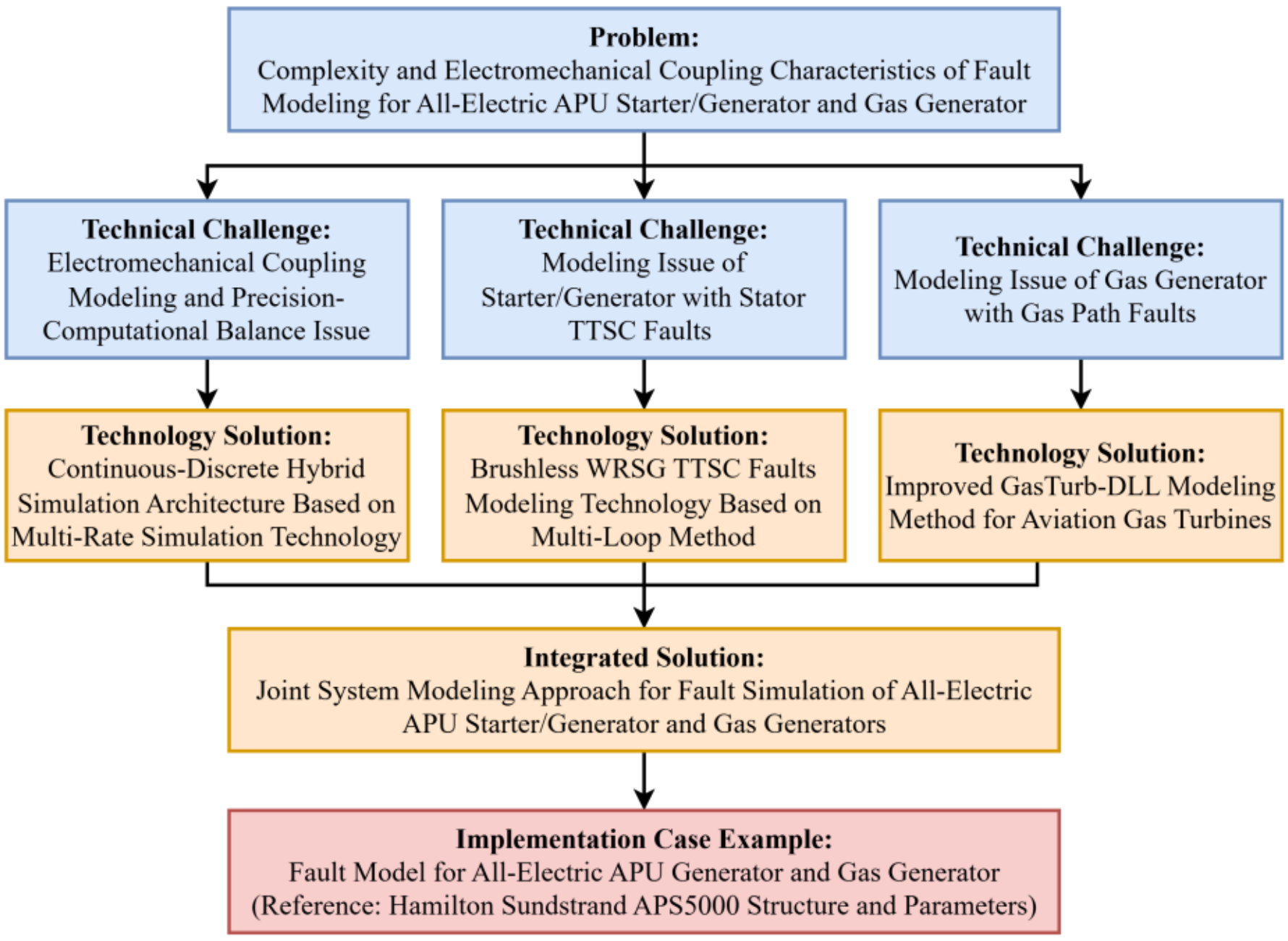

**Figure 3.** Research framework (problem, work, technical innovations, and achievements)

To address these challenges, this paper proposes a joint system modeling approach for fault simulation of all-electric APU starter/generators and gas generators. This approach implements joint system modeling for fault simulation of complex electromechanical systems through three technical innovations: (1) a multi-rate continuous-discrete hybrid simulation architecture that effectively solves the electromechanical coupling modeling challenge while significantly reducing computational load without compromising accuracy; (2) multi-loop electrical machine modeling technology for simulating both healthy and TTSC fault conditions in brushless wound rotor synchronous starter/generator (WRSG); and (3) an improved GasTurb-DLL modeling method for aviation gas turbines that offers advantages in uncertainty modeling and simulation, state-space representation, and tool chain compatibility.

Finally, using these methods and techniques, a joint system model for fault simulation of the all-electric APU starter/generator and gas generator was developed based on the Hamilton Sundstrand APS5000 all-electric APU structure and parameters. The effectiveness of both the joint system modeling approach and the established all-electric APU model was verified through extensive simulations and comparison with third-party data.

## 3. Multi-Rate Continuous-Discrete Hybrid Simulation Architecture

To optimize computational efficiency while ensuring joint model accuracy, this research proposes a multi-rate continuous-discrete hybrid all-electric APU model architecture, as shown in Figure 4. This architecture adopts a mixed modeling strategy with different simulation rates and solving environments, modeling the starter/generator system as a variable-step continuous system in the Simulink environment, while modeling the gas generator system as a fixed-step discrete system in the dynamic-link library (DLL) environment, achieving precise matching of system characteristics and rational allocation of computational resources.

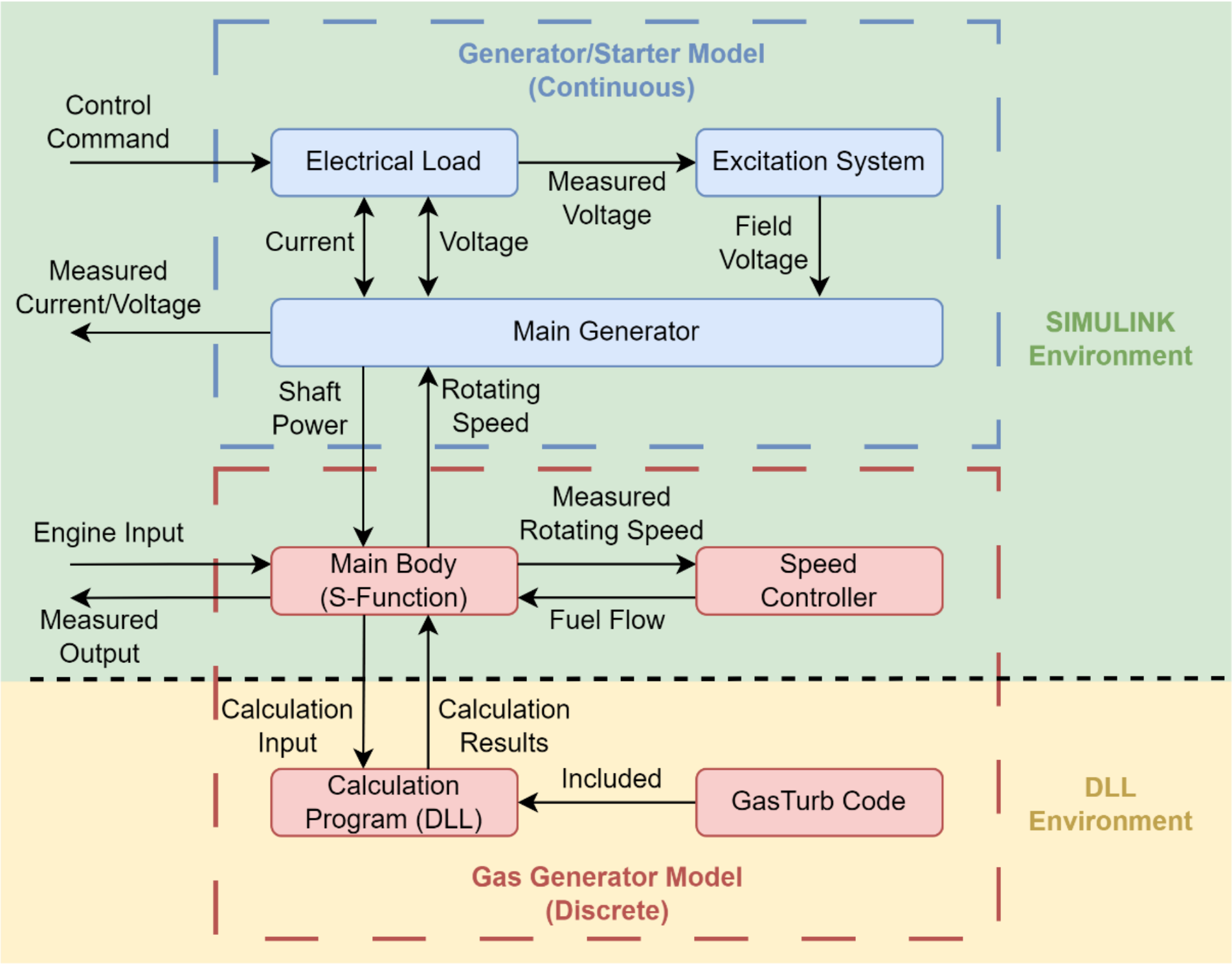

**Figure 4.** Proposed Multi-Rate Continuous-Discrete Hybrid Simulation Architecture of All-Electric APU Starter/Generator and Gas Generator

Multi-rate simulation, as an efficient simulation method, has been widely applied in electromechanical systems such as wind turbines and ground gas turbines [39-42]. These systems typically comprise electronic and mechanical subsystems, where electronic systems feature high-frequency, fast-response characteristics requiring shorter simulation steps to ensure accuracy, while mechanical systems exhibit large inertia and low response frequencies that can be satisfied with longer simulation steps. Multi-rate simulation applies differentiated step sizes to different subsystems based on their characteristics and response frequencies, significantly reducing computational cost while ensuring simulation accuracy, achieving an optimal balance between computational efficiency and simulation precision [40-43].

The hybrid simulation architecture integrates different modeling environments and solvers to model power systems and mechanical systems respectively, utilizing the advantages of each environment and solver to achieve efficient, accurate, and reliable system joint system modeling.

Next, we will elaborate on our proposed hybrid simulation architecture in the following sequence: first, designing a multi-rate hybrid simulation strategy for all-electric APU; second, completing the overall design of the hybrid simulation architecture, including model structure construction and sub-module layout optimization; and finally, formulating effective parameter synchronization and transfer strategies to facilitate proper information exchange between systems operating at different simulation rates, thereby minimizing hybrid simulation errors and enhancing overall model simulation accuracy.

### *3.1. Hybrid Simulation Strategy Design*

As shown in Figure 4, the all-electric APU model constructed in this research consists of two core modules: the starter/generator model and the gas generator model. Based on system characteristic analysis, this paper employs the following multi-rate hybrid simulation strategy: variable-step continuous model for the starter/generator and fixed-step discrete model for the gas generator. This design decision is based on the following technical considerations.

The decision to use a variable-step continuous model for the starter/generator is primarily based on the following key factors: First, the starter/generator involves complex electromagnetic calculations with a high dynamic response frequency range. Using traditional discrete fixed-step solvers would lead to exponential growth in computational load. Second, when current and voltage fluctuate dramatically, such as during faults, fixed-step solvers struggle to ensure calculation accuracy. Continuous variable-step solving methods can adaptively adjust step size according to system dynamic characteristics, effectively controlling computational load while ensuring calculation accuracy [44-46].

For the gas generator model, the fixed-step discrete model design is based on the following considerations: Gas turbine systems have larger mechanical inertia and lower dynamic response frequencies, with complex thermodynamic calculation processes. Using a continuous model would introduce unnecessary computational overhead. By adopting a fixed-step discrete model, computational complexity can be significantly reduced while maintaining calculation accuracy [16,37,47-49].

Based on the above hybrid simulation strategy, the following solving environments and solvers were selected. As shown in Figure 4, the entire model adopts a dual-environment parallel solving architecture: SIMULINK environment and DLL environment. The SIMULINK environment is responsible for solving the starter/generator model, gas generator S-Function module, and speed controller; the DLL environment is specifically used for solving the gas generator aerodynamic thermodynamic cycle program.

The advantage of this dual-environment architecture design lies in its ability to leverage the characteristics of each platform to build an efficient, reliable, and high-precision all-electric APU model. Specifically, the DLL environment can directly call GasTurb's Pascal language code, utilizing GasTurb's component library to build the aviation gas turbine model. In this environment, GasTurb's built-in Newton-Raphson-based solver can be used to solve the gas generator model, a solver renowned in the industry for its efficiency, reliability, and precision, ensuring model accuracy and computational speed [25,37,47,48,50].

In the SIMULINK environment, the electrical system and control system component libraries and various solvers in its Simscape system can be utilized to efficiently build engine models and gas generator controllers. Additionally, Matlab's ode23tb solver can be used for numerical calculation of the starter/generator model. As a high-performance stiff solver, ode23tb exhibits excellent stability and reliability when handling continuous models of electrical systems. This solver can not only process continuous system calculations but also accommodate discrete gas generator model calculation results, enabling joint modeling of electrical systems and gas generator systems [51-54].

The above describes the multi-rate hybrid simulation strategy and corresponding dual-environment solution structure proposed in this research. To avoid conceptual confusion, we provide further explanation of the continuous and discrete concepts here. The continuity or discreteness of a model refers to the mathematical equation forms employed during the modeling process: continuous models typically describe system dynamics using differential equations, capable of precisely capturing continuous state changes; while discrete models express system behavior at discrete time points through difference equations or state transition functions. In the actual numerical solution process, regardless of whether the model is continuous or discrete, discretization methods are ultimately required for computation. However, different solution strategies and solvers can be selected according to the model characteristics, as outlined in the design principles discussed earlier.

### *3.2. Overall Simulation Architecture Design*

Next, we will introduce the details of the architecture, as shown in Figure 4. The architecture contains a starter/generator model and a gas generator model. The starter/generator model includes a main generator, excitation system, and power load. The aircraft controls the power load through control signals, and the power load is connected to the starter/generator, influencing each other and forming dynamic voltage and current that are measured and output by sensors. These measured values are obtained by the excitation system to control the excitation voltage output by the excitation system, thereby keeping the output voltage amplitude of the main generator within the specified range. All-electric APU generally install two starter/generators; to simplify the problem, this paper does not consider the working coordination and scheduling between the two starter/generators and assumes that their working states, fault degrees, loads, etc., are completely consistent, so only one starter/generator is modeled, and its power is doubled to represent the total power of both starter/generators.

The gas generator part consists of two components: the gas generator's main body module, which calculates various mechanical and aerothermodynamic output parameters, as well as speed parameters required by the starter/generator module based on the shaft power input from the generator and other engine inputs; and a speed controller used to maintain constant speed, which controls engine speed through fuel flow and speed feedback.

The main body module is an S-function module whose main function is to define the attributes and inputs/outputs of the main body module, and during simulation, to obtain inputs from the Simulink environment, call the aerothermodynamic calculation program for calculation, and finally output the results back to the Simulink environment.

The aerothermodynamic program calculation program is called in DLL form. This calculation program itself utilizes GasTurb-related functions and code, including GasTurb solver code and component-level aerothermodynamic calculation code for aviation gas turbines. Specific gas generator modeling details will be further elaborated in Section 5.

### 3.3. Parameter Synchronization and Transfer Strategy

The all-electric APU multi-rate continuous model-discrete hybrid modeling method adopted in this research can adapt to the characteristics of the generator and gas generator, using different simulation step sizes and simulation strategies to achieve the goal of ensuring accuracy while greatly reducing the computational load. However, this method introduces the problem of asynchronous parameter updates between discrete and continuous models, which may lead to certain model errors. Therefore, a set of parameter synchronization and transfer rules for this all-electric APU hybrid model was designed to minimize hybrid simulation errors and improve the overall model simulation accuracy.

As shown in Figure 4, the parameters that need to be synchronized and transferred between the starter/generator and the gas generator include shaft speed and shaft power. For these two parameters, to solve the problem of inconsistent update step sizes between the starter/generator model and the gas generator model, the following parameter transfer rules were designed:

$$Pe^{gt}(t_k) = \frac{\int_{t_{k-1}}^{t_k} P^{sg-total}(t)dt}{\left(t_k - t_{k-1}\right)\eta^{gtTsg}} \quad (1)$$

$$w_r^{sg}(t) = \frac{w_r^{gt}(t_k)}{\Omega^{gtTsg}(t_k)} \quad (2)$$

where $P^{sg-total}$ represents the total power input of the starter/generator, $Pe^{gt}$ represents the shaft power output by the gas generator, $\eta^{gtTsg}$ represents the power transfer efficiency from the gas generator to the starter/generator, $w_r^{sg}$ represents the starter/generator rotational speed, $w_r^{gt}$ represents the output shaft rotational speed of the gas generator, and $\Omega^{gtTsg}$ represents the speed ratio between the gas generator and starter generator determined by the gearbox.

Compared to traditional APU, all-electric APU have eliminated constant speed drives (CSD) and instead employ variable frequency power system. Consequently, in all-electric APU, the rotational speed ratio between the starter/generator and the gas generator is maintained at a fixed value determined by the gearbox [55]. The speed transfer between the gas generator and starter/generator in equations (1) and (2) introduce a fixed speed ratio $\Omega^{gtTsg}$, and power parameter transfer also needs to introduce transmission efficiency $\eta^{gtTsg}$.

Next, it will be explained that under this parameter transfer processing method, there is no error when the starter/generator transfers parameters to the gas generator, and the error when the gas generator transfers parameters to the starter/generator is very small and can be ignored.

First is the process of transferring power parameters from the starter/generator to the gas generator. Under the above parameter transfer rules, for any second, the shaft power output value required for gas generator calculation can be obtained from equation (1). Therefore, the power balance equation of the gas generator can be derived as follows:

$$\begin{aligned}\varepsilon^{gt-power}(t_k) &= [P^{gt-turb}(t_k) - \frac{P^{gt-cpr}(t_k)}{\eta^{cpr}(t_k)} - Pe^{gt}(t_k)](t_k - t_{k-1}) \\ &= [P^{gt-turb}(t_k) - \frac{P^{gt-cpr}(t_k)}{\eta^{cpr}(t_k)}](t_k - t_{k-1}) - \int_{t_{k-1}}^{t_k} \frac{P^{sg-total}(t)}{\eta^{gtTsg}(t_k)} dt\end{aligned} \quad (3)$$

where $\varepsilon^{gt-power}$ represents the residual of the power balance equation, $\eta^{cpr}$ represents the power transfer efficiency of the compressor, $P^{gt-cpr}$ represents the power input of the compressor, and $P^{gt-turb}$ represents the power output of the turbine.

From the derivation process in equation (3), it can be seen that when calculating the power balance of the gas generator, the gas generator power output per second equals the input power of the synchronous motor plus the power transmission losses during the corresponding time. Therefore, there is no error due to different parameter update frequencies when the gas generator model performs calculations.

Next is the process of transferring speed parameters from the gas generator to the starter/generator. The simulation step length of the gas generator model constructed in this paper is set to 20ms, which is commonly used in the field of aviation gas turbine fault diagnosis [47,49]. Since the APU speed does not fluctuate over a wide range except during startup, but rather revolves around the working speed, and the gas generator itself is limited by the inertia of the rotor system, the system response characteristics are slow. Therefore, within the selected simulation step, the gas turbine speed remains almost constant. Hence, when using equation (2) to transfer speed from the gas generator to the starter/generator, the error can be ignored.

The above describes the designed multi-rate simulation continuous-discrete hybrid simulation architecture, which achieves efficient joint modeling and solving of the starter/generator and gas generator through multi-rate, multi-environment simulation, ensuring accuracy while minimizing computational load as much as possible.

## 4. Starter/Generator Model with TTSC Faults

This section details the modeling approach, assumptions, and implementation process for the all-electric APU starter/generator. First, the structure of the starter/generator, modeling principles, and underlying assumptions are introduced. Then, the multi-loop modeling approach for handling stator TTSC faults is explained in detail, along with the key equations governing the starter/generator system dynamics, and how the model is implemented and solved in the MATLAB/SIMULINK environment. Finally, a detailed structural diagram of the starter/generator model is presented, providing a comprehensive summary of the model.

### *4.1. Modeling Assumptions*

This paper takes the brushless wound rotor synchronous starter/generator (WRSG), widely used in both all-electric APU and traditional APU, as the modeling object, establishing a continuous model of all-electric APU starter/generator TTSC faults that can run in the MATLAB/SIMULINK environment.

The brushless WRSG, whose structure is shown in Figure 5 [56], consisting of four parts: permanent magnet generator, main exciter, main generator, and electrical load. The first three parts are the main structure, while the last one, the electrical load, is adjusted according to different simulation tasks to simulate different system working conditions.

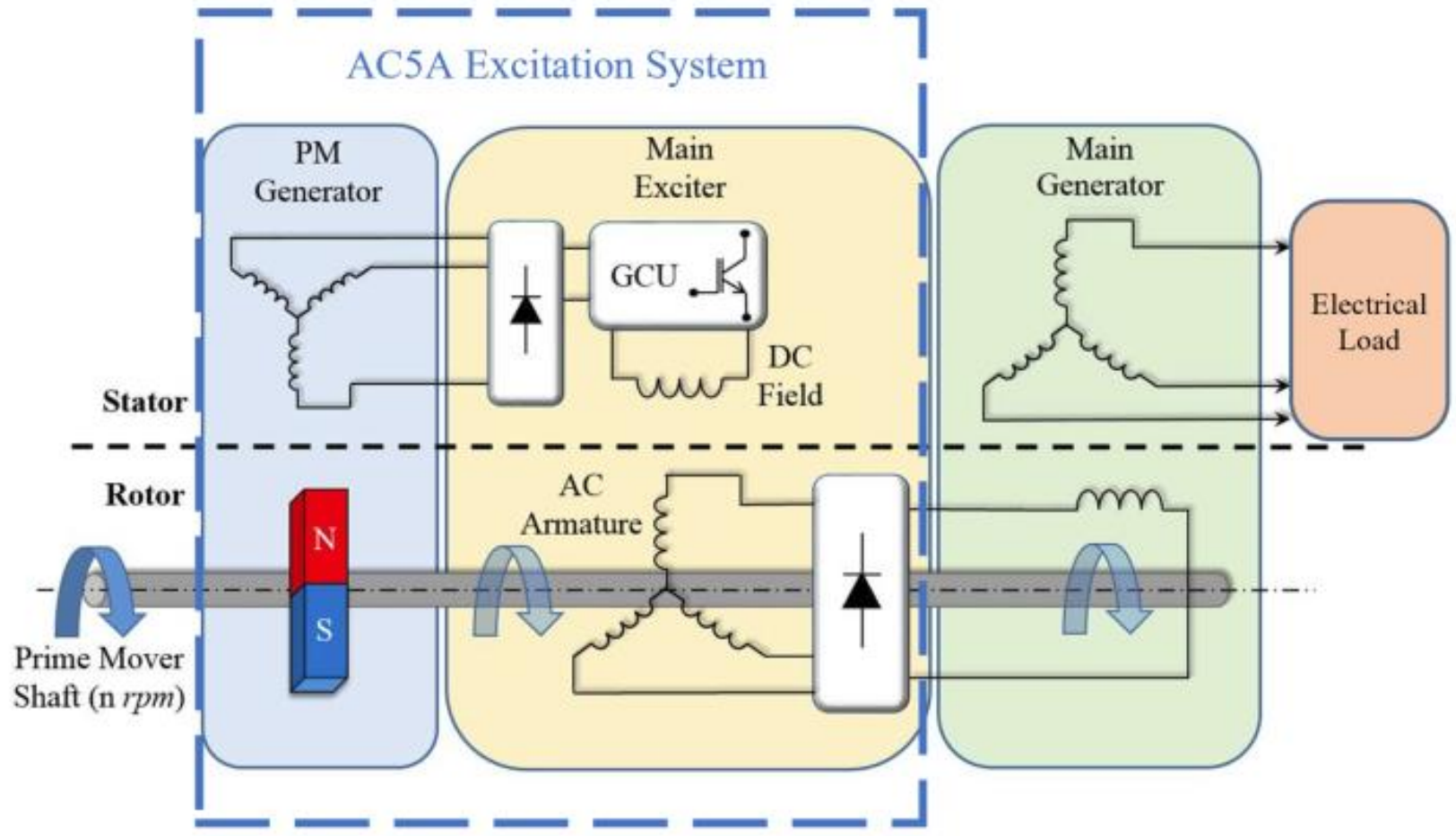

**Figure 5.** Schematic diagram of the brushless WRSG

For the first two parts of the starter/generator—the permanent magnet synchronous machine and the main exciter—a simplified approach is taken. Since this paper focuses specifically on the stator winding TTSC faults in the main generator part, the detailed structure and parameters of the excitation system are not examined in depth. Instead, these two components are simplified and modeled as an IEEE AC5A excitation system [31,57,58], as shown in Figure 5. Furthermore, this study assumes that the stator TTSC faults is in its early stages with relatively small magnitude, and that the starter/generator has sufficient design redundancy to allow us to ignore magnetic saturation effects [59].

For the third part of the starter/generator, the main generator part, it is assumed that its three-phase stators are sa, sb, sc, outputting three-phase alternating current, with the TTSC fault occurring in phase sa, as shown in Figure 6. It is also assumed that the fault branch resistance of the faulty sa phase is $R_f$, the fault current is $i_f$, the direction is as shown in the figure, and the fault causes the number of shorted turns to account for the ratio of the total number of stator coil turns. The shorted part of the coil in phase sa is $\mu$, the healthy part of the coil is $sa-f$, and the positive direction of each phase current is $sa-h$ as shown in the figure.

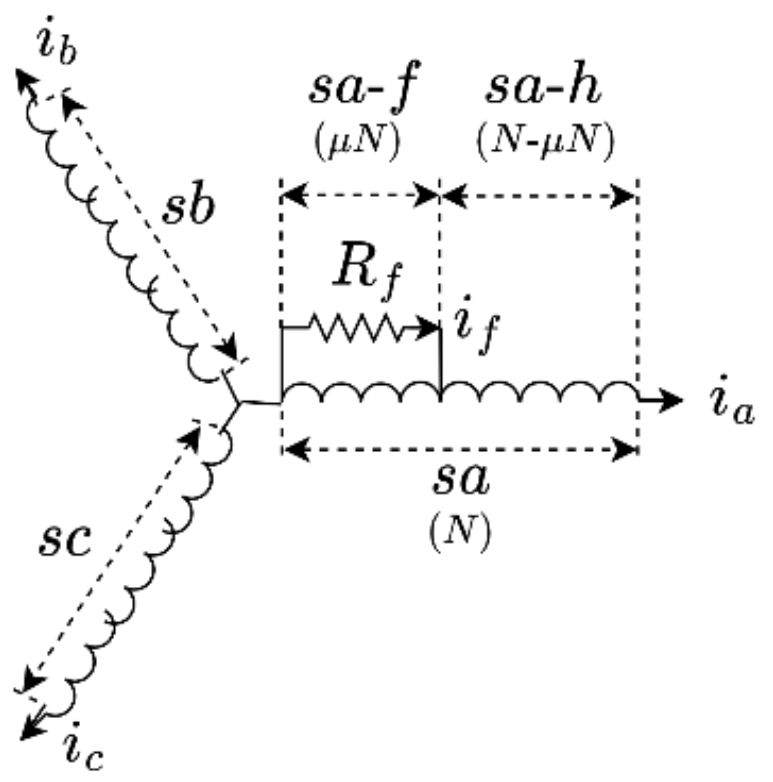

**Figure 6.** Schematic diagram of stator TTSC faults

To better handle the problem, the parameter $k_{rf}$ is introduced, representing the ratio of the short circuit branch resistance to the resistance of the shorted coil, so the short circuit branch resistance can be expressed by the following equation (4). Thus, the stator TTSC fault situation can be described by two fault parameters $k_{rf}$ and $\mu$.

$$r_f = k_{rf} \mu r_s \tag{4}$$

Next, the main generator is modeled by using a widely adopted and validated multi-loop theory [59-61]. This approach assumes that the TTSC fault results in the formation of a new current path, as depicted in Figure 6, causing a diversion of current, and thus reducing the current flowing through the original turns. During this process, the number of turns in the faulty phase does not decrease. Based on this assumption, the multi-loop theory treats the short-circuit branch formed during a TTSC fault as a fault loop. Subsequently, by computing the coupling effects of the fault current in this loop with other parts of the generator, the operational state of the generator under this fault condition can be deduced. This methodology enables the representation of both healthy scenarios (where the short-circuit branch resistance is near-infinite) and faulty scenarios (where the short-circuit branch resistance is relatively low).

In addition, the dq0 reference system is used to reduce the complexity of the inductance matrix used to describe the relationship between flux linkage and current, and to reduce the required computational resources. The orientation of coordinate axes and the directionality of various quantities used in this research refer to the IEEE standards [62]. Next, the key equations used to determine the main generator system dynamics will be presented. Since the starter/generator model adopts a continuous model, continuous equations will be used for description, and time symbols will be omitted.

### 4.2. Modeling Method based on Muli-loop theory

The relationship between the voltage, current, and flux linkage of the armature winding of the main generator can be described by

$$V_{s-qd0} = W_r \lambda_{s-qd0} + \frac{d\lambda_{s-qd0}}{dt} - TR_s T^{-1} i_{s-qd0} + T\mu r_s i_{f-abc} + w_1 \tag{5}$$

where

$$V_{s-qd0} = [V_{sq}, V_{sd}, V_{s0}]^T,\ \lambda_{s-qd0} = [\lambda_{sq}, \lambda_{sd}, \lambda_{s0}]^T,\ i_{s-qd0} = [i_{sq}, i_{sd}, i_{s0}]^T$$

$$i_{f-abc} = [i_f, 0, 0]^T,\ R_s = diag[r_s, r_s, r_s]^T,\ W_r = \begin{bmatrix} 0 & w_r & 0 \\ -w_r & 0 & 0 \\ 0 & 0 & 0 \end{bmatrix}$$

$V_{sq}, V_{sd}, V_{s0}$ : Denote the voltage of the stator winding in the dq0 rotating reference frame

$\lambda_{sq}, \lambda_{sd}, \lambda_{s0}$: Denote the flux linkage of the stator winding in the dq0 rotating reference frame

$i_{sq}, i_{sd}, i_{s0}$: Denote the current of the stator winding in the dq0 rotating reference frame

$T$ : Denotes the transformation matrix from the abc frame to the dq0 frame

$r_s$ : Denotes the Stator phase winding resistance

$w_r$ : Denotes the Generator speed, and

$w_1$ : Denotes the Noise indicating uncertainty.

The rotor of the main generator can be classified into two types, namely salient and round. Our research focuses on the salient-pole rotor used in all-electric APU starter/generator for subsequent investigation. Then, the relationship between the voltage, current, and flux linkage of the rotor winding in the generator is described by

$$V_{r-qd0} = \frac{d\lambda_{r-qd0}}{dt} + R_r i_{r-qd0} + w_2 \tag{6}$$

where

$$\lambda_{r-qd0} = [\lambda_{fd}, \lambda_{kd}, \lambda_{kq}]^T,\ V_{r-qd0} = [V_{fd}, V_{kd}, V_{kq}]^T$$

$i_{r-qd0} = [i_{fd}, i_{kd}, i_{kq}]^T$, $R_r = diag[r_{fd}, r_{kd}, r_{kq}]^T$

$V_{fd}, V_{kd}, V_{kq}$ : Denote the voltage of the rotor field winding, damper bars on the d-axis and q-axis

$\lambda_{fd}, \lambda_{kd}, \lambda_{kq}$ : Denote the flux linkage of rotor field winding, damper bars on the d-axis and q-axis

$i_{fd}, i_{kd}, i_{kq}$ : Denote the current of rotor field winding, damper bars on the d-axis and q-axis

$r_{fd}, r_{kd}, r_{kq}$ : Denote the resistance of rotor field winding, damper bars on the d-axis and q-axis, and

$w_2$ : Denotes the Noise indicating uncertainty.

The relationship between the flux linkage and current in the starter/generator can be expressed by

$$\lambda_{s,r-qd0} = L(i_{s,r-qd0} - K_1 T_{(:,1)} \mu i_f) \quad (7)$$

where

$i_{s,r-qd0} = [i_{sq}, i_{sd}, i_{s0}, i_{fd}, i_{kd}, i_{kq}]^T$, $\lambda_{s,r-qd0} = [\lambda_{sq}, \lambda_{sd}, \lambda_{s0}, \lambda_{fd}, \lambda_{kd}, \lambda_{kq}]^T$

$$K_1 = \begin{bmatrix} 1 & 0 & 0 & 0 & 0 & 0 \\ 0 & 1 & 0 & 0 & 0 & 0 \\ 0 & 0 & 1 & 0 & 0 & 0 \end{bmatrix}^T$$

$T_{(:,1)}$ : The first column of the matrix T, and

$$L = \begin{bmatrix} -(L_{mq} + L_{ls}) & 0 & 0 & 0 & 0 & L_{mq} \\ 0 & -(L_{md} + L_{ls}) & 0 & L_{md} & L_{md} & 0 \\ 0 & 0 & -L_s & 0 & 0 & 0 \\ 0 & -L_{md} & 0 & (L_{md} + L_{lfd}) & L_{md} & 0 \\ 0 & -L_{md} & 0 & L_{md} & (L_{md} + L_{lkd}) & 0 \\ -L_{mq} & 0 & 0 & 0 & 0 & (L_{md} + L_{lkq}) \end{bmatrix}.$$

The above equations (5)-(7) determine the voltage-current relationship of the generator. It should be noted that, as this study focuses on the initial stages of the TTSC malfunction, the inductance matrix does not take into account the effects of distortion in the space harmonics, air-gap flux density disturbance, among others that are caused by the TTSC [28,59,63]. However, our theory is applicable to any form of the inductance matrix, and hence is capable of being extended. This suggests that, if necessary, more accurate results that consider the effects of distortion and air-gap flux density disturbances can be obtained by integrating other analytical methods, such as the winding function approach, to refine the inductance matrix [31].

Based on voltage and current, it is easy to calculate the output power of the starter/generator and the power loss caused by the TTSC. Combined with the starter/generator efficiency, the total mechanical power $P^{sg-total}$ of the two starter/generator can be expressed by equation (8) [64].

$$P^{sg-total} = 2\frac{V_{abc} i_{abc}^T + i_f^2 r_f + (i_a - i_f)^2 r_{sa-f} - i_a^2 r_{sa-f}}{\eta_{sg}} \quad (8)$$

Due to presence of stator TTSC faults, the above equations contain parameters that are related to the fault. To investigate the above equation, it is necessary to introduce the following equations to describe the fault circuit, namely

$$\frac{d\lambda_{sa-f}}{dt} - \mu r_s (i_a - i_f) = -R_f i_f \quad (9)$$

where $\lambda_{sa-f}$ denotes the flux linkage of the faulty winding $sa-f$ in phase $sa$.

The generator terminal measurements of voltage and current can be expressed using equation (10). Equations (5)-(10) collectively determine the current-voltage characteristics of the main generator. Based on these equations, we can establish the required main generator model. This model, together with the aforementioned excitation system and the specific load (which varies according to different tasks), will constitute the complete starter/generator model capable of simulating stator inter-turn short circuits. The specific structure, modeling details, and parameters of this model will be presented in the following section.

$$\begin{bmatrix} \tilde{i}_{s-abc} \\ \tilde{V}_{s-abc,fd} \end{bmatrix} = \begin{bmatrix} i_{s-abc} \\ V_{s-abc,fd} \end{bmatrix} + \begin{bmatrix} v_i \\ v_v \end{bmatrix} \tag{10}$$

where

$V_{s-abc,fd} = [V_{sa}, V_{sb}, V_{sc}, V_{fd}]^T$, $i_{s-abc} = [i_{sa}, i_{sb}, i_{sc}]^T$

$V_{sa}, V_{sb}, V_{sc}$ : Denote the terminal voltage of the three-phase

$i_{sa}, i_{sb}, i_{sc}$ : Denote the terminal current of the three-phase

$v_i, v_v$ : Denote the uncertainty associated with current and voltage measurements, respectively, and

~: The superscript of the measured values.

### *4.2. Implementation of the Starter/Generator Model*

Based on the theoretical framework outlined above, a starter/generator model with TTSC fault injection capability was developed in the MATLAB/SIMULINK environment. The structural details of this model are illustrated in Figure 7. The parameters of the brushless WRSG are derived from reference [65] and the Hamilton Sundstrand APS5000 all-electric APU, as listed in Table 1.

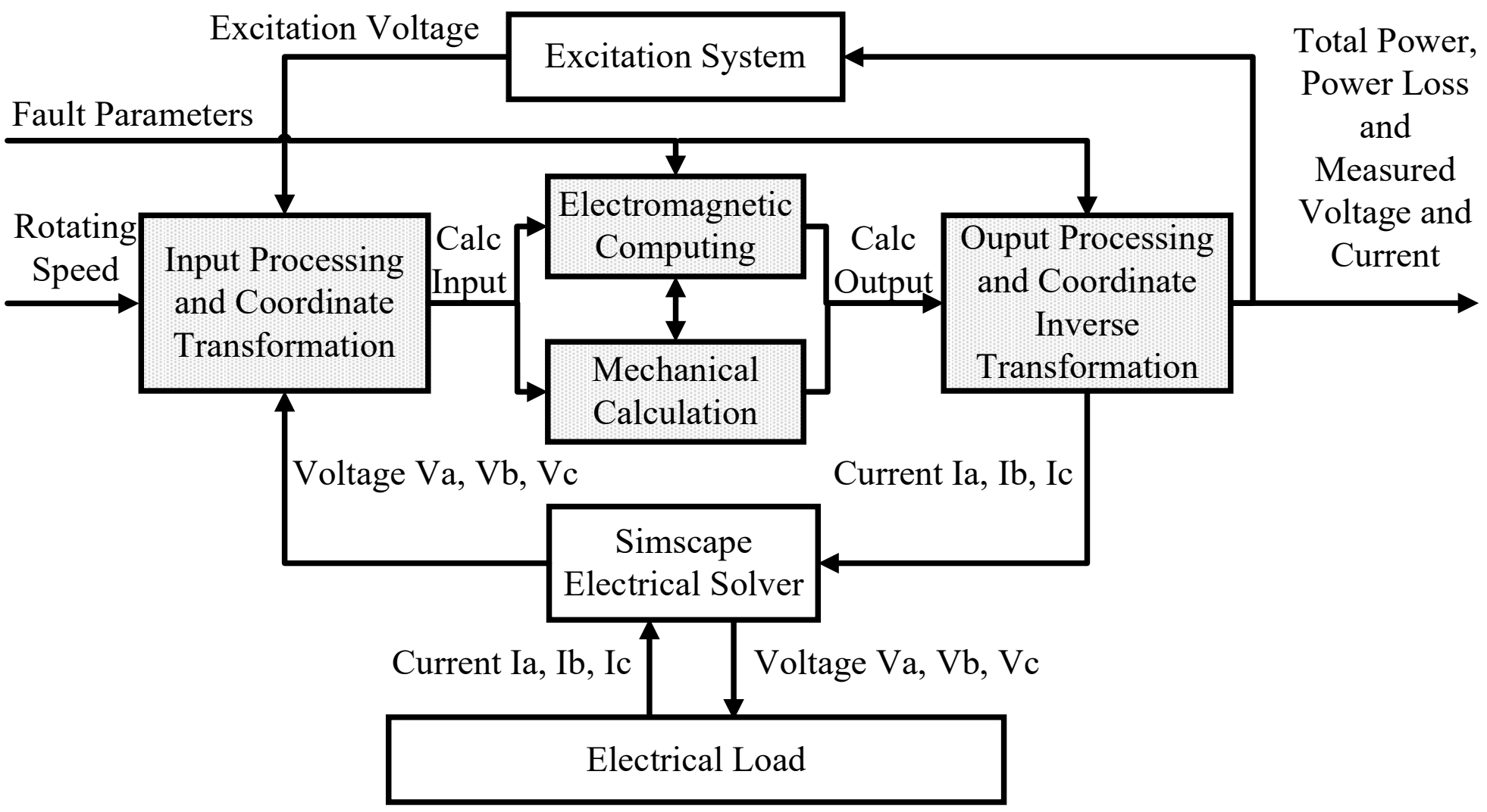

**Figure 7.** Structural diagram of the starter/generator model

In the starter/generator model structure depicted in Figure 7, the four dark gray modules constitute the main generator component. Initially, the input processing and coordinate system transformation module is responsible for preprocessing, unit conversion, and abc-dq0 coordinate transformation of the input signals, including the rotational speed $w_r$ from gas generator, the excitation voltage $V_{fd}$ from the excitation system, and the terminal voltage $V_{s-abc}$ from the solver. Subsequently, the processed parameters are transmitted to both the electromagnetic calculation module and the mechanical calculation module. The electromagnetic calculation module calculates electromagnetic parameters such as flux linkages $\lambda_{s,r-qd0}$, $\lambda_{sa-f}$, and implements TTSC faults based on fault parameters $\mu$, $k_{rf}$. The mechanical calculation module computes mechanical parameters such as rotor angle. Finally, the calculation results and fault parameters are forwarded to the output processing and coordinate transformation module, which handles output preprocessing, unit conversion, and dq0-abc coordinate transformation. This module outputs the calculated terminal current $i_{s-abc}$ to the solver, and output the total power $P^{sg-total}$ to the gas generator, the power loss $P^{sg-loss}$ due to TTSC faults, and the measured terminal voltage and current $\tilde{V}_{s-abc}$, $\tilde{i}_{s-abc}$ for FDI and data validation.

**Table 1.** The set of starter/generator parameters

| Parameters | Description | Unit | Values |
|---|---|---|---|
| $P_n$ | Rated Power | kW | 225 |
| $V_{rms}$ | Line voltage | V | 230 |
| $f_n$ | Frequency | Hz | 400 |
| $R_s$ | Stator resistance | Ω | 0.0044 |
| $L_{ls}$ | Stator leakage inductance | μH | 19.8 |
| $L_{md}$ | Stator d-axis magnetizing inductance | mH | 0.221 |
| $L_{mq}$ | Stator q-axis magnetizing inductance | mH | 0.162 |
| $R_f$ | Field resistance | mΩ | 68.9 |
| $L_{lf}$ | Field leakage inductance | μH | 32.8 |
| $R_{kd}$ | d-axis resistance | Ω | 0.0142 |
| $L_{lkd}$ | d-axis leakage inductance | μH | 34.1 |
| $R_{kq}$ | q-axis resistance | Ω | 0.0031 |
| $R_{lkq}$ | q-axis leakage inductance | mH | 0.144 |
| $P$ | Pole pairs of the generator | / | 2 |

In this process, the excitation system shown in Figure 7 collects the starter/generator's output voltage signals $\tilde{V}_{s-abc}$ and adjusts the excitation voltage $V_{fd}$ accordingly to maintain proper output voltage. The electrical load system requires specific configuration based on different simulation requirements. The Simscape power solver employs the aforementioned ode23tb algorithm to iteratively solve for the terminal voltage Vt and terminal current It of the entire starter/generator model system.

## 5. Gas Generator Fault Model

This section details the modeling approach, assumptions for the gas generator, and proposes an improved GasTurb-DLL modeling method (IGDM) based on which the required gas generator model was implemented.

First, the section introduces the modeling principles, and assumptions for the gas generator; then, it details the traditional GasTurb-DLL modeling method (GDM) [37,48,66] and its limitations; next, it proposes the IGDM method and explains in detail how this method improves upon the GDM method to resolve its limitations; finally, it showcases the gas generator model developed based on the proposed IGDM method and its design parameters.

### *5.1. Modeling Assumptions*

This paper takes the single-shaft turboshaft engine used in mainstream all-electric APU as the modeling object, establishing a discrete model of the all-electric APU gas generator with speed controller that can simulate gas path faults and run in the MATLAB/SIMULINK environment.

The gas generator gas path fault model mainly consists of a gas generator aerothermodynamic model and a speed controller. Since the focus of this paper is not engine control, the controller used is a simple PI controller that uses speed feedback to control fuel. As for the gas generator body, it is essentially an aviation gas turbine, where a single-shaft turboshaft engine is chosen as the prototype, which from front to back includes a compressor, combustion chamber, turbine, and exhaust pipe. It should be noted that this type of turboshaft configuration is also widely adopted in contemporary all-electric APU. The overall structure, section numbering, and cooling airflow direction are shown in Figure 8 and Table 2 [67,68].

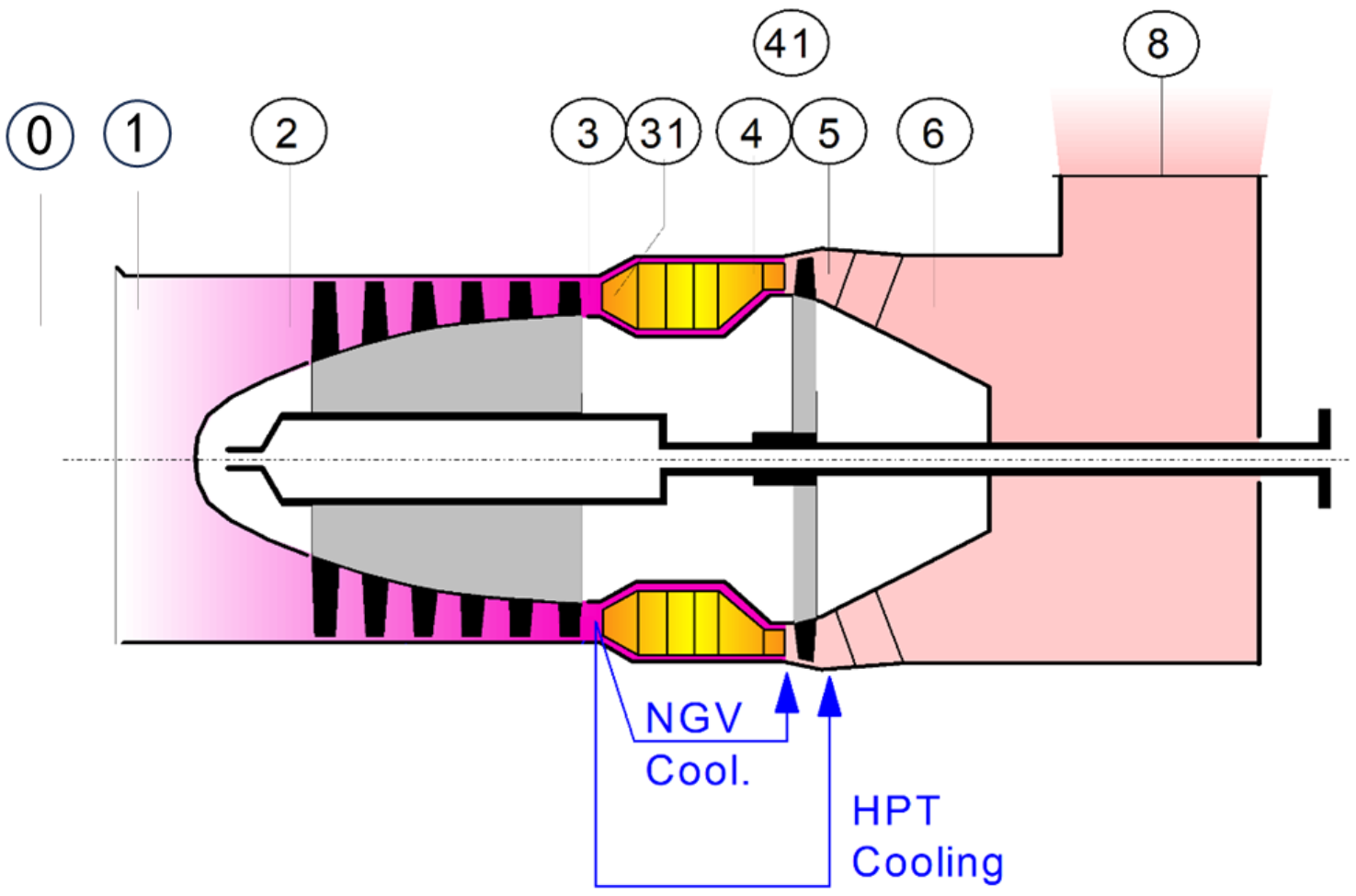

**Figure 8.** Structural diagram of the gas generator model

**Table 2.** Cross-section Definitions of Single-shaft Turboshaft Gas Generator

| Number | Definition Description | Number | Definition Description |
|---|---|---|---|
| 0 | Atmosphere | 1 | Air Intake Entrance |
| 2 | Compressor Inlet | 3 | Compressor Outlet |
| 31 | Combustion Chamber Inlet | 4 | Combustion Chamber Outlet |
| 41 | Turbine Inlet | 5 | Turbine Outlet |
| 6 | Exhaust System Inlet | 8 | Exhaust System Outlet |

### *5.2. GDM Method and Its Limitations*

In this section, the modeling principles and processes of the GDM method [37,48,66] will first be introduced, followed by an in-depth analysis and discussion of the limitations of the GDM method in terms of tool chain compatibility, uncertainty modeling and simulation of systems, as well as the state space representation of systems.

#### 5.2.1. GDM Method

The GDM method builds aeronautical gas turbine models through a systematic approach. Initially, it utilizes the widely recognized GasTurb software to generate a preliminary model, which is stored in GasTurb file format. Subsequently, through MATLAB/SIMULINK's S-function modules and DLL technology, the model is configured to operate in a modular form within the MATLAB/SIMULINK environment, resulting in the desired gas turbine model module. The details will be elaborated in the subsequent discussion.

In the GDM method, a gas turbine model is first established in GasTurb software based on specific requirements (structure, performance, key parameters), and exported as a configuration file. This file defines the aerothermodynamic design parameters and performance characteristics of the gas turbine.

The core of the GDM method involves three Pascal-language functions: TransitionInitialize, TransitionState, and TransitionFree. These functions leverage GasTurb's aerothermodynamic calculation capabilities to compute the gas turbine's parameters and dynamics based on inputs at each timestep, the designed parameters, and structure (from the configuration file).

The TransitionInitialize function handles environment initialization and calculates the initial steady-state conditions at t=0s. Its inputs include the gas turbine configuration file, simulation parameters, and initial inputs. After initialization, it computes the steady state parameters and relevant aerothermodynamic values corresponding to the initial inputs, saves these as the initial state, and outputs selected parameters.

The TransitionState function calculates the state parameters and aerothermodynamic parameters for each timestep after t=0s. Using the current inputs and previous state, it computes the current state parameters and outputs the selected parameters.

The TransitionFree function terminates calculation and releases allocated memory.

These three functions constitute the core aerothermodynamic calculation process in the GDM method. The workflow, illustrated in Figure 9, begins with TransitionInitialize for initialization and initial steady-state output, followed by repeated calls to TransitionState for dynamic outputs at each subsequent timestep, and concludes with TransitionFree. These functions are compiled into a 32-bit DLL (GT10DLL.dll) for subsequent S-function implementation.

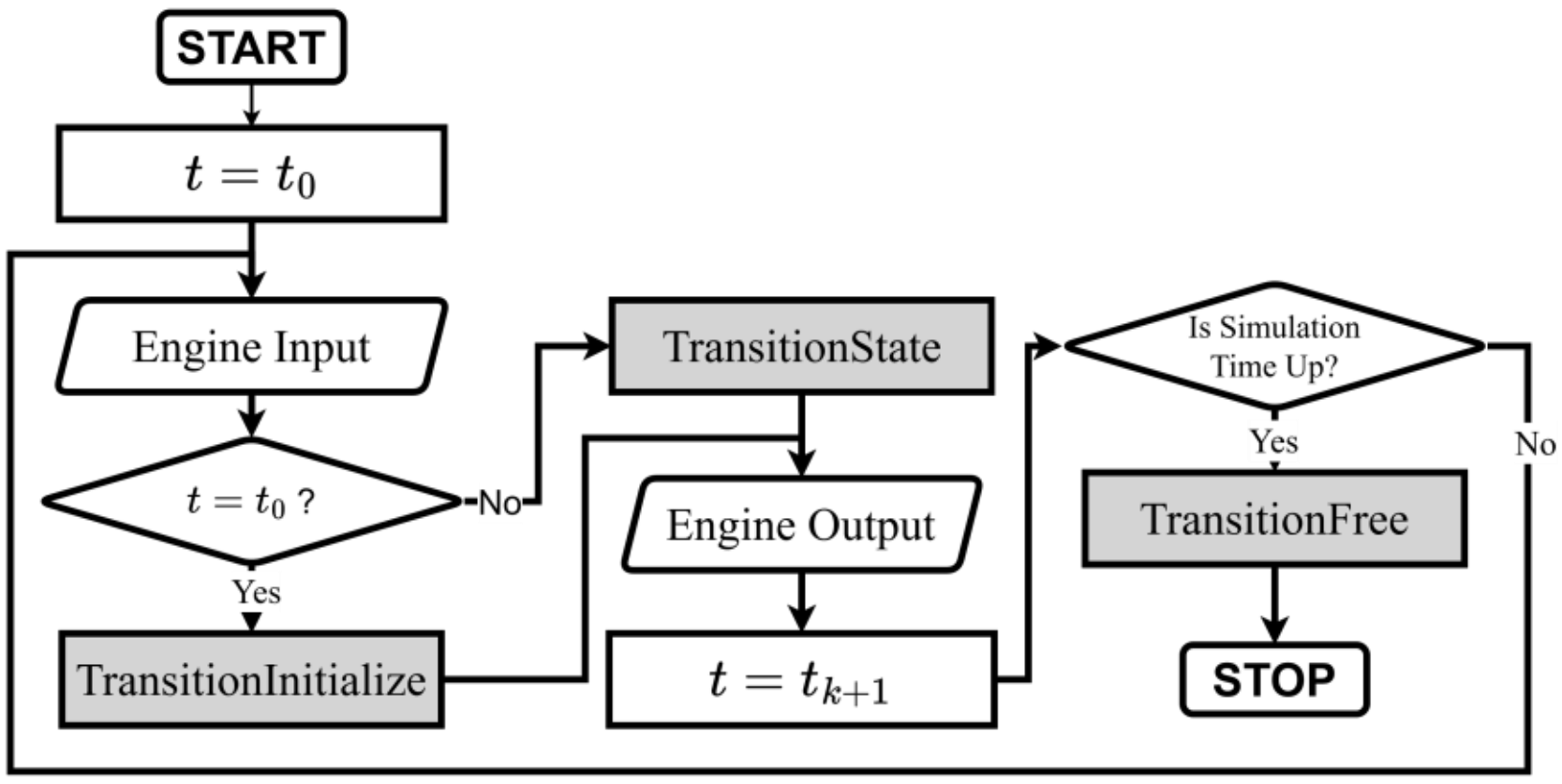

**Figure 9.** Aerothermodynamic calculation flow of the GDM method model

After encapsulating these functions as a DLL, the GDM method employs MATLAB/SIMULINK's C++-based S-function to call these functions and package them into a modular form. This requires developing a C++ program called "civilengine," compiled into mexw32 format for MATLAB/SIMULINK. This S-function program defines initialization, input/output handling, and termination procedures according to MATLAB/SIMULINK standards. During execution, it retrieves module inputs, calls the appropriate DLL functions, and formats the results as module outputs.

The complete GDM model generation process is illustrated in Figure 10. The process begins with determining the gas turbine model's requirements, establishing the model in GasTurb, and developing necessary functions. The configuration file is exported from GasTurb, the S-function source code is compiled, and the aerothermodynamic functions are exported as a DLL. Finally, using MATLAB/SIMULINK's S-function module and Mask functionality, the complete gas turbine module for gas generator is created.

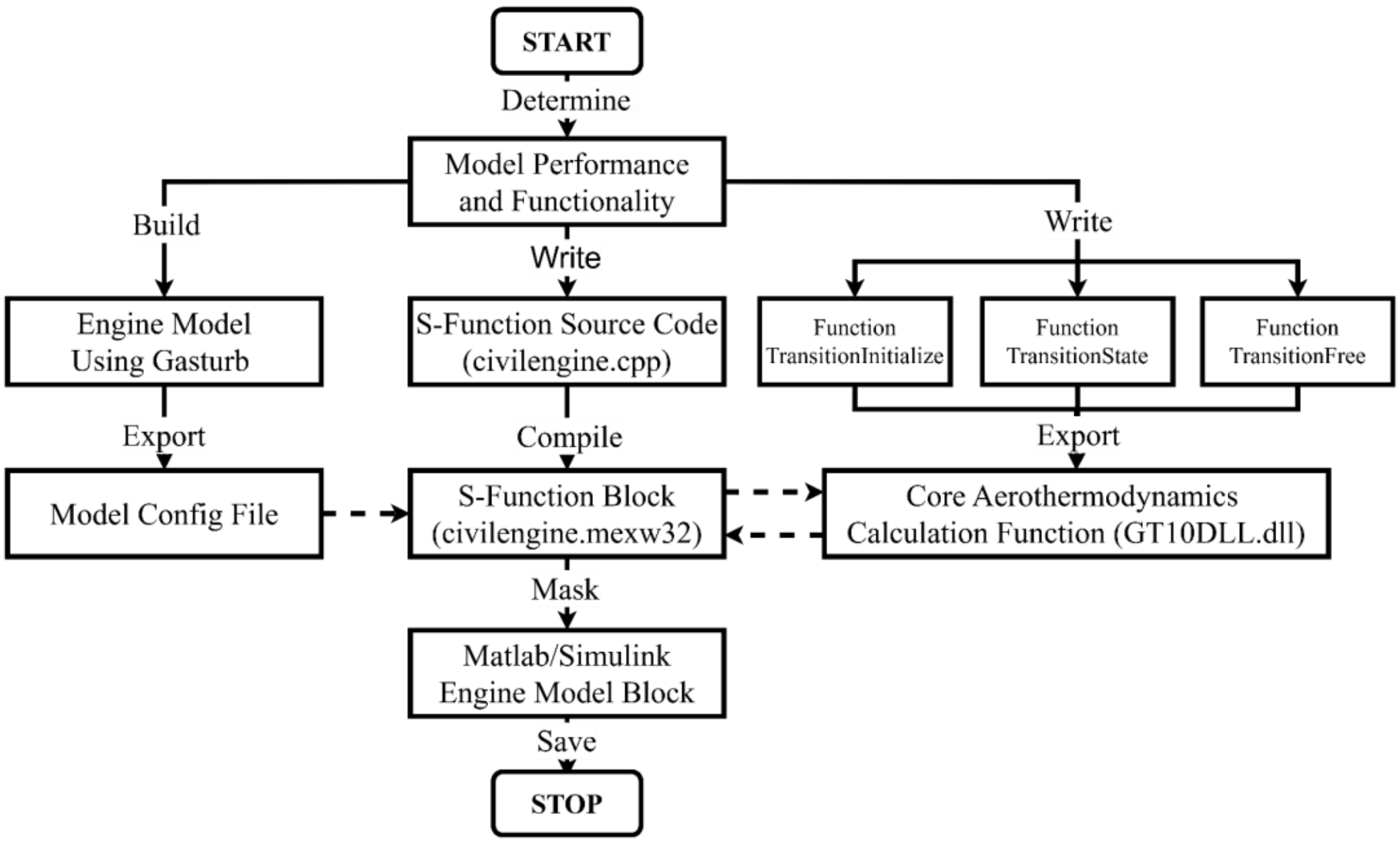

**Figure 10.** Flow chart of GDM method generating aviation gas turbine model

5.2.2. Limitations of the GDM Method

Despite its effectiveness, the GDM method exhibits two significant limitations: its 32-bit architecture and its implicit state updating mechanism. These limitations impact tool chain compatibility, uncertainty modeling and simulation, and state-space representation of the generated model.

1. 32-bit Architecture Limitations

The GDM method produces 32-bit models because the TransitionInitialize, TransitionState, and TransitionFree functions must call GasTurb functions written for 32-bit platforms. Consequently, the "civilengine" S-function program must be compiled as a 32-bit binary (mexw32). Since 64-bit MATLAB cannot call mexw32 programs, and Windows restricts 64-bit processes from using 32-bit DLLs, the GDM-generated models can only run in 32-bit MATLAB environments.

As MATLAB has not released 32-bit versions since 2015b, GDM models are restricted to this decade-old version, which lacks many features and performance improvements of current MATLAB releases. This significantly constrains the toolchain and technical solutions available for research based on these models, affecting areas such as power system modeling, neural network applications, simulation analysis tools, and hardware-in-the-loop simulations.

2. Implicit State Updating Limitations

The iteration algorithm of GDM-generated models, shown in Figure 11, reveals that the model's mathematical form can be expressed as:

$$y(t_{0:k}) = f_{GDM}\left(u(t_{0:k}), \theta(t_{0:k}), Pe(t_{0:k})\right) \quad (11)$$

where $f_{GDM}$ represents the GDM model calculation program, $y$ represents the model output, $u$ represents the model input, $\theta$ represents the gas path health parameters describing faults and degradation, and $Pe$ represents the shaft power of the load.

**Algorithm 1:** Iterative Algorithm for GDM

**Input** : Engine input: $u(t_k)$;
Gas health parameter: $\theta(t_k)$;
Electric generator power requirements $P(t_k)$;
**Output:** Engine Output: $y(t_k)$;

**if** $k = 0$ **then**
  `/* Initialization */`
  $y(t_k) \leftarrow TransitionInitialize(u(t_k), \theta(t_k), P(t_k))$;
**else**
  `/* Output Calculation and Update */`
  $y(t_k) \leftarrow TransitionState(u(t_k), \theta(t_k), P(t_k))$;
**end**
$k = k + 1$;
**if** $k > LastTimeStep$ **then**
  `/* End of Calculation */`
  $TransitionFree()$;
**end**

**Figure 11.** GDM method iteration algorithm

This formulation demonstrates that state variables are not explicit inputs to the GDM model. Although state variables can be included in the output $y$ (such as shaft speed), their updating process remains implicit and uncontrollable. This means that while the model can simulate the gas turbine and compute outputs from inputs, the state updating and output processes are coupled, yielding only the final output results without access to the intermediate state updating process.

This coupling prevents expressing the gas turbine in standard state-space form and precludes adding noise to the state update equations to introduce state uncertainty. For fault diagnosis and prediction studies, introducing uncertainties is essential to represent differences between models and reality. The common approach adds state noise and output noise to the system's state update and output equations [16,69-71]. The GDM method's implicit state updating allows only output noise, not state noise, limiting the representation of model-reality uncertainty [16,69,70].

Furthermore, the implicit state updating prevents controlled simulation of given system states for Monte Carlo simulation purposes, further limiting uncertainty modeling and simulation capabilities.

### *5.3. IGDM Method: Improved GDM Modeling Method*

To address the limitations of the GDM method, we developed IGDM. Compared to GDM, IGDM incorporates the following enhancements: (1) modification of the model code for 64-bit compatibility, thereby eliminating toolchain compatibility issues, and (2) restructuring of the state updating process to make it explicit and controllable, with state updates separated from output processes. These improvements enable the addition of state noise, facilitate Monte Carlo simulation of state variables, and provide standard state-space representation.

Specifically, IGDM achieves these improvements and addresses GDM limitations through enhancements to the aerothermodynamic core function development and export, as well as through S-function development and compilation processes.

#### 5.3.1. Improvements to Aerothermodynamic Core Functions

The IGDM method develops four Pascal-language functions for gas turbine aerothermodynamic calculations: TransitionInitialize_I, TransitionStateUpdate_I, TransitionOutput_I, and TransitionFree_I. These correspond to and enhance the GDM functions, with TransitionStateUpdate_I and TransitionOutput_I together replacing the original TransitionState function.

First, transitionInitialize_I improves upon TransitionInitialize in two ways: it explicitly outputs state variables and achieves 32-bit and 64-bit compatibility. By modifying the output code, it explicitly outputs the initial steady-state values to provide initial state information for the subsequent TransitionStateUpdate_I function. It also modifies both its own code and the GasTurb functions it calls to ensure compatibility with both 32-bit and 64-bit environments.

Second, the most significant improvement is the decomposition of TransitionState into TransitionStateUpdate_I and TransitionOutput_I. This separation accomplishes two critical objectives:

- It transforms implicit state updating into explicit updating, separating the state update process from the output process, and allows the addition of state update noise.
- It achieves 32-bit and 64-bit compatibility through code modifications.

In the original GDM method, TransitionState calculated the current gas turbine state and aerothermodynamic parameters based on inputs and the previous state. The state updating occurred internally within the function, preventing external control or modification of the process. This limitation made it impossible to add state noise or perform controlled state Monte Carlo simulations.

By splitting this function into state update (TransitionStateUpdate_I) and output (TransitionOutput_I) components, IGDM transforms state variables from implicit to explicit controllable variables. TransitionStateUpdate_I implements the state update equation, explicitly outputting the next state, while TransitionOutput_I implements the output equation, taking the explicit state as input to calculate the current gas turbine outputs. This separation provides a standard state-space representation and allows the addition of both state and output noise, enabling comprehensive uncertainty modeling.

Third, TransitionFree_I, like the other functions, maintains compatibility with both 32-bit and 64-bit environments while performing the same memory release function as the original TransitionFree.

The resulting aerothermodynamic calculation process is illustrated in Figure 12. Compared to the GDM workflow as shown in Figure 9, the key improvement is the decomposition of TransitionState into separate state update and output functions, providing an explicit state update control interface.

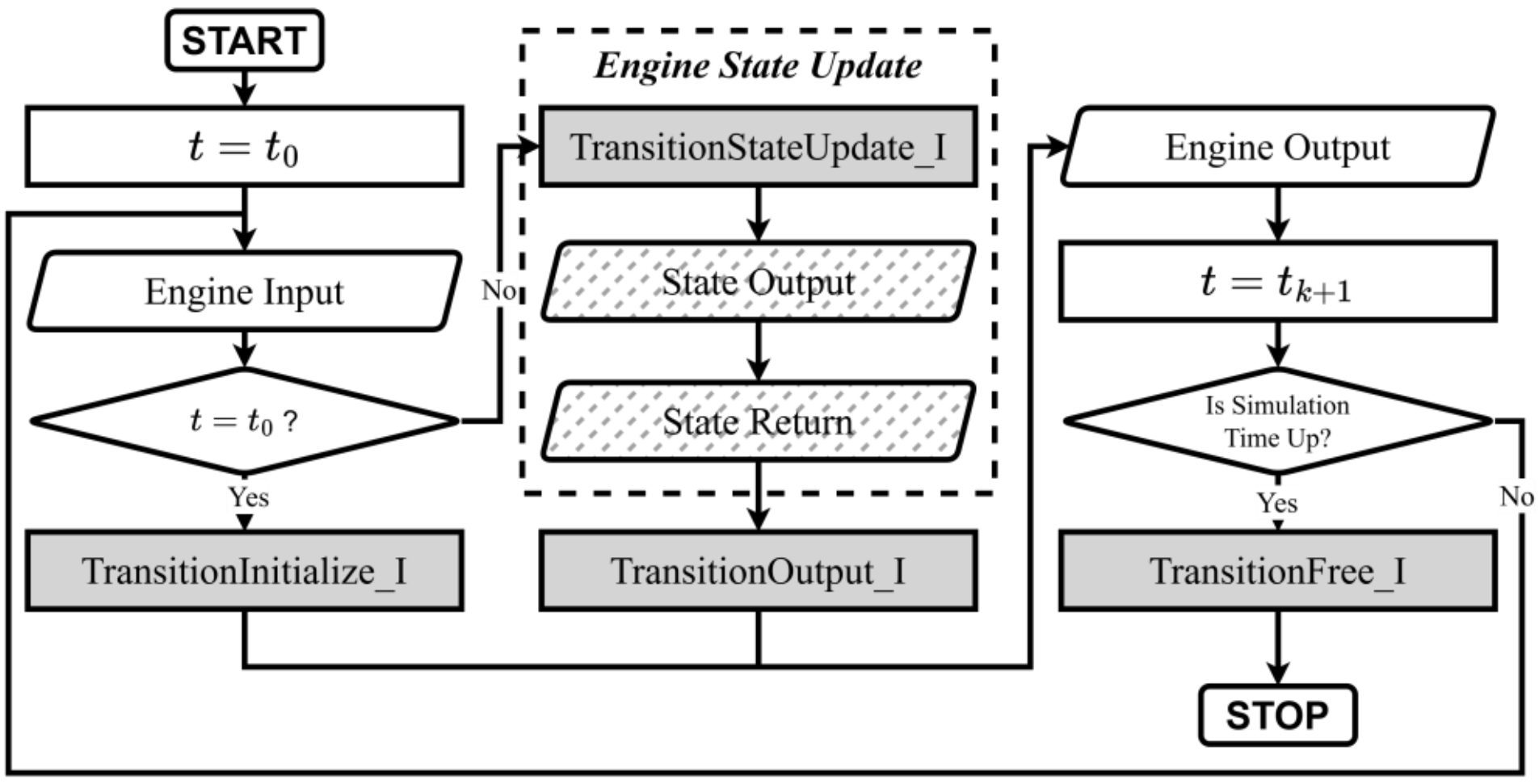

**Figure 12.** Aerothermodynamic calculation flow of the IGDM method model

From an algorithmic perspective, as shown in Figure 13, IGDM implements explicit state control through the StateProcess interface function, which controls interaction with TransitionStateUpdate_I and TransitionOutput_I.

**Algorithm 2:** Iterative Algorithm for IGDM

**Input :** Engine input $u(t_k)$
Gas path health parameter $\theta(t_k)$
Electric generator power requirements $P(t_k)$
Engine Speed $N_L(t_k)$
**Output:** Engine Output $y(t_k)$

$x(t_k) = N_L(t_{k-1})$;
**if** $k = 0$ **then**
  /* Initialization */
  $y(t_k) \leftarrow TransitionInitialize_I(u(t_k), \theta(t_k), P(t_k))$;
**else**
  /* State Calculation */
  $x(t_{k+1}) \leftarrow TransitionStateUpdate_I(u(t_k), x(t_k), \theta(t_k), P(t_k))$ ;
  /* State output, externally processed, returned, updated */
  $x(t_{k+1}) = StateProcess(x(t_{k+1}))$;
  /* Output Calculation and Update */
  $y(t_k) \leftarrow TransitionOutput_I(u(t_k), x(t_k), \theta(t_k), P(t_k), x(t_k)) + v(t_k)$;
**end**
$k = k + 1$;
**if** $k > LastTimeStep$ **then**
  /* End of Calculation */
  $TransitionFree()$;
**end**

**Figure 13.** IGDM method iteration algorithm

These improvements enable IGDM to address GDM's limitations in uncertainty modeling and simulation: the model can incorporate state noise, perform controlled Monte Carlo simulations of different system states, and express the system in standard state-space form:

$$\begin{aligned} x(t_k) &= f_{IGDM}\left(x(t_{k-1}), u(t_k), \theta(t_k), Pe(t_k)\right) \\ y(t_k) &= g_{IGDM}\left(x(t_k), u(t_k), \theta(t_k), Pe(t_k)\right) \end{aligned} \tag{12}$$

where $f_{IGDM}$ and $g_{IGDM}$ represent IGDM model calculation programs, $y$ represents the model output, $u$ represents the model input, $x$ represents the system state, $\theta$ represents the gas path health parameters, and $Pe$ represents the shaft power of the load.

After completing these functions, they are packaged in Delphi IDE and exported as both 32-bit and 64-bit DLLs: GT10DLL_I32.dll and GT10DLL_I64.dll.

### 5.3.2. Improvements to S-function Development

Following the DLL encapsulation, IGDM employs MATLAB/SIMULINK's C++-based S-function capability to call these functions and create modular gas turbine models. Two versions of the "civilengine" program are developed: civilengine_I32 calling GT10DLL_I32.dll and civilengine_I64 calling GT10DLL_I64.dll. These are compiled into mexw32 and mexw64 formats respectively, for use in both 32-bit and 64-bit MATLAB/SIMULINK environments.

### 5.3.3. IGDM Method Summary

The IGDM method for developing MATLAB/SIMULINK gas turbine models is illustrated in Figure 14. Like GDM, it comprises three major components: GasTurb model development and configuration export, S-Function development and compilation, and aerothermodynamic core function development and DLL compilation. The key differences lie in the S-function and aerothermodynamic core function development.

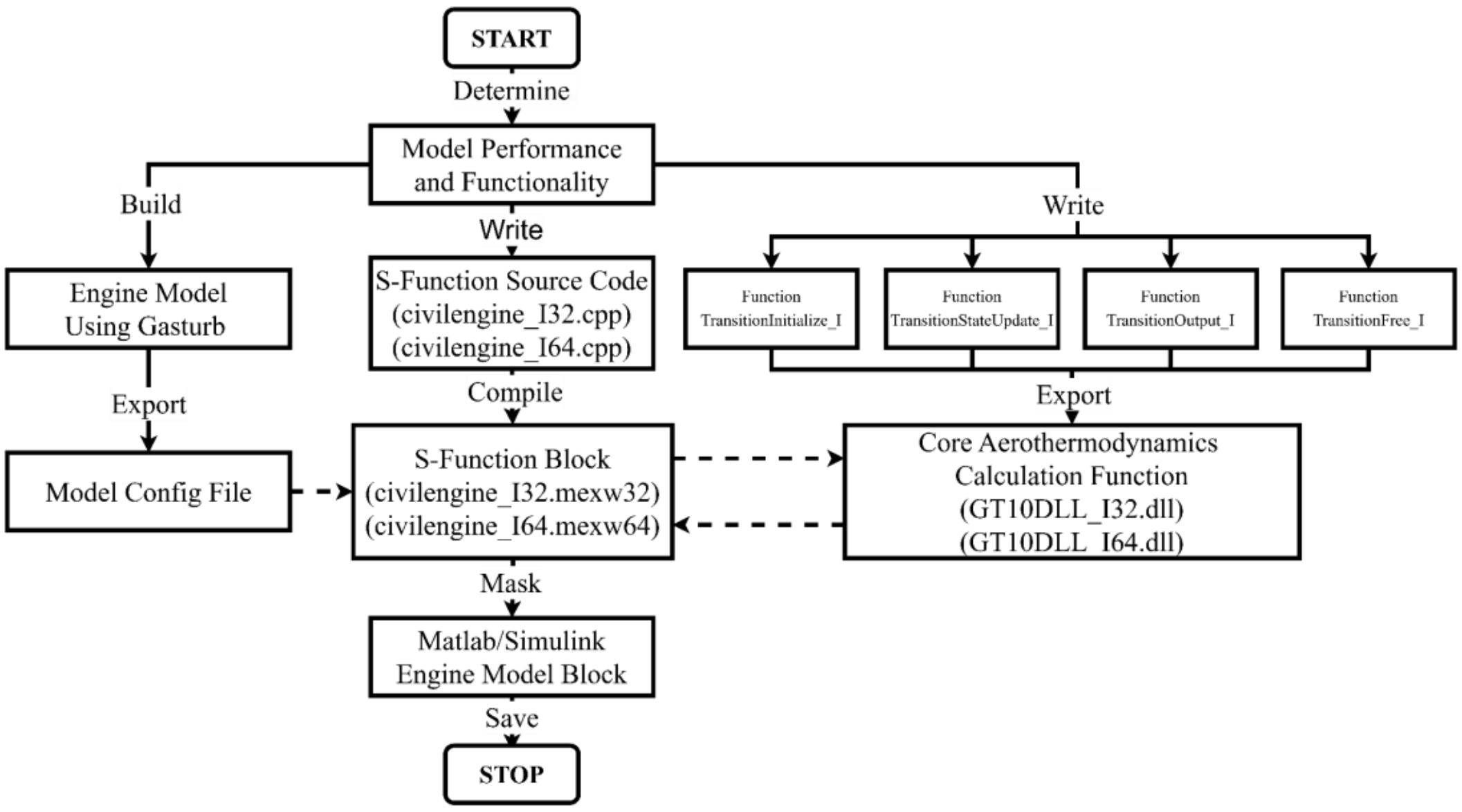

**Figure 14.** Flow chart of IGDM method generating aviation gas turbine model

IGDM achieves 32-bit and 64-bit compatibility throughout the S-function and aerothermodynamic function development, compilation, and export processes. This enables IGDM models to run in both newer and older versions of MATLAB/SIMULINK, resolving GDM's toolchain compatibility limitations.

Additionally, by separating the state update process from the output process, IGDM provides explicit, controllable state-space updating. This enables the addition of state noise, facilitates Monte Carlo simulation of state variables, and allows representation in standard state-space form, addressing GDM's limitations in uncertainty modeling, simulation, and state-space representation.

### *5.4. Implementation of the Gas Generator Model*

Next, based on the IGDM method proposed above, a single-shaft turboshaft engine will be established as the main body part of gas generator, combined with the previously mentioned speed controller, to construct the required gas generator model module. During modeling, the engine's combustion delay is not considered, changes in generation efficiency and shaft power transfer efficiency are not considered, the engine system state variable is selected as the rotor speed, and compressor and turbine efficiency factors and flow factors are introduced to inject and describe gas path faults [16-20].

Referring to the gas generator parameters of the Hamilton Sundstrand APS5000 all-electric APU, the gas generator is specifically designed and configured, with its main design parameters shown in the following Table 3.

According to the IGDM method process, the above parameter configuration and design are completed in GasTurb, the S-function code and dynamic link library code required for modeling are written, and using these codes, the required single-shaft turboshaft engine module is built in the MATLAB/SIMULINK environment as the gas generator main body. Connecting this main body module with the aforementioned speed controller completes the construction of the required gas generator simulation model that can simulate gas path faults.

**Table 3.** The set of gas generator parameters

| Parameter | Unit | Value |
|---|---|---|
| Altitude | m | 0 |
| Mach Number | / | 0 |
| ISA Deviation | K | 5 |
| Design Point Output Shaft Power | kW | 500 |
| Compressor Pressure Ratio | / | 8 |
| Combustion Chamber Outlet Temperature | K | 1200 |
| Exhaust Temperature | K | 755 |
| Fuel Heating Value (Fuel Type: Generic) | MJ/kg | 43.124 |
| Compressor Inlet Corrected Flow | kg/s | 2.383 |
| Compressor Efficiency | / | 0.85 |
| Design Speed | r/min | 36050 |
| Turbine Efficiency | / | 0.89 |
| Accessorial Power | kW | 30 |

## 6. Simulation Results

This section will simulate the established joint system model for fault simulation of starter/generator and gas generator in all-electric APU (SG-EG Model), and verify the effectiveness of the model through GasTurb software, MATLAB synchronous motor module (SM-MATLAB), and data from relevant literature, thus demonstrating that the constructed model can correctly simulate and model the healthy and fault working conditions of the gas generator and starter/generator, providing support for all-electric APU FDI research.

The verification process will first involve module-level simulation targeting the main components of the joint simulation fault model—specifically the gas generator module (EG module) and the starter/generator module (SG module)—to validate their effectiveness. Following this, the entire joint system model will be simulated and verified to ensure its overall performance and accuracy.

### *6.1. Gas Generator Model Simulation and Verification*

First, for the gas generator module, single-point steady state simulation at the design point and multi-point steady simulation at off-design points will be conducted, and the results will be verified using GasTurb software; next, the transients and faults of the gas generator module will be simulated, and the results will also be verified using GasTurb software.

#### 6.1.1. Design Point and Off-Design Point Steady State Simulation and Verification for Gas Generator

The design point of all-electric APU refers to the specific operating condition at which the all-electric APU determines its primary design parameters and performance. Off-design points, on the other hand, encompass the various stable operating conditions of the all-electric APU outside of this design point.

In the subsequent research, first, for the design point of the gas generator module, sea level, 20°C, 500kW shaft power, simulation is conducted, and at the same time, using GasTurb software, a simulation is conducted for an engine model with the same design parameters, design point operating conditions and fuel flow (0.04830 kg/s). The simulation results of both, comparison, and errors are shown in the following Table 4.

**Table 4.** Design point: 0km 0Ma 500kW

| Parameter | Description | Unit | GasTurb | EG Module | Relative Error |
|---|---|---|---|---|---|
| XNHPC | Rotor Speed | r/min | 36050.0000 | 36050.0039 | $1.0836\times10^{-7}$ |
| PWSD | Output Shaft Power | kW | 500.0027 | 499.9999 | $5.6152\times10^{-6}$ |
| SFC | Specific Fuel Consumption | kg/(kW.h) | 0.3280 | 0.3280 | $5.2693\times10^{-6}$ |
| SNOx | NOx Severity Factor | / | 0.1769 | 0.1769 | $1.1795\times10^{-6}$ |
| HPCSM | Compressor Stability Margin | / | 23.9856 | 23.9855 | $3.8170\times10^{-6}$ |
| T1 | Inlet Total Temperature | K | 293.1500 | 293.1500 | $1.3573\times10^{-15}$ |
| P1 | Inlet Total Pressure | kPa | 101.3250 | 101.3250 | $1.8233\times10^{-15}$ |
| T2 | Compressor Inlet Total Temperature | K | 293.1500 | 293.1500 | $1.3573\times10^{-15}$ |
| P2 | Compressor Inlet Total Pressure | kPa | 100.3118 | 100.3118 | $4.6750\times10^{-15}$ |
| W2 | Compressor Inlet Flow Rate | kg/s | 3.1442 | 3.1442 | $7.5829\times10^{-8}$ |
| T3 | Compressor Outlet Total Temperature | K | 568.1270 | 568.1270 | $4.0022\times10^{-16}$ |
| P3 | Compressor Outlet Total Pressure | kPa | 802.4929 | 802.4935 | $8.3663\times10^{-7}$ |
| Ps3 | Compressor Outlet Static Pressure | kPa | 768.3194 | 768.3201 | $8.7384\times10^{-7}$ |
| W3 | Compressor Outlet Flow Rate | kg/s | 3.1442 | 3.1442 | $7.5829\times10^{-8}$ |
| T4 | Combustion Chamber Outlet Total Temperature | K | 1200.1140 | 1200.1140 | $4.1681\times10^{-15}$ |
| P4 | Combustion Chamber Outlet Total Pressure | kPa | 778.4180 | 778.4187 | $8.6250\times10^{-7}$ |
| W4 | Combustion Chamber Outlet Flow Rate | kg/s | 2.8466 | 2.8466 | $1.6751\times10^{-7}$ |
| T41 | Turbine Inlet Total Temperature | K | 1169.3879 | 1169.3879 | $4.2776\times10^{-15}$ |
| W41 | Turbine Inlet Flow Rate | kg/s | 3.0038 | 3.0038 | $1.5874\times10^{-7}$ |
| T5 | Turbine Outlet Total Temperature | K | 755.1480 | 755.1480 | $3.0110\times10^{-16}$ |
| P5 | Turbine Outlet Total Pressure | kPa | 106.4943 | 106.4946 | $3.0806\times10^{-6}$ |
| W5 | Turbine Outlet Flow Rate | kg/s | 3.1610 | 3.1610 | $1.5085\times10^{-7}$ |
| T8 | Engine Outlet Total Temperature | K | 755.1480 | 755.1480 | $3.0110\times10^{-16}$ |
| P8 | Engine Outlet Total Pressure | kPa | 104.3644 | 104.3647 | $3.2897\times10^{-6}$ |
| W8 | Engine Outlet Flow Rate | kg/s | 3.1610 | 3.1610 | $1.5085\times10^{-7}$ |
| Average Relative Error：$1.0285\times10^{-6}$ | | | | | |

From the table, it can be seen that the design point simulation calculation results of the constructed gas generator module are almost identical to the GasTurb calculation results, with average relative error of $1.0285\times10^{-6}$, maximum relative error of $5.6152\times10^{-6}$ for the output shaft power PWSD, and minimum relative error of $3.0110\times10^{-16}$ for T5 and T8. Therefore, this verifies the design point effectiveness of the gas generator module.

Next, for the off-design points of the gas generator module, by adjusting the altitude, Mach number, and shaft power, 10 different operating conditions were selected for simulation, and at the same time, using GasTurb software, simulation was conducted for engine models with the same design parameters and operating conditions. Due to space limitations, only the simulation results for one high-altitude 222kW shaft power condition are shown, as in Table 5.

From Table 5, it can be seen that the calculation results of the constructed gas generator model and GasTurb have very small errors, with an average relative error of only $1.1021\times10^{-6}$. And under all 10 selected off-design point conditions, the average error between the two parameters is $1.3461\times10^{-5}$, slightly higher than the design point, but still very small. Therefore, this verifies the off-design point effectiveness of the gas generator module.

**Table 5.** Off-design point: 8000km 0.7Ma 222kW

| Parameter | Description | Unit | GasTurb | EG Module | Relative Error |
|---|---|---|---|---|---|
| XNHPC | Rotor Speed | r/min | 36050.0000 | 36050.0508 | $1.4086\times10^{-6}$ |
| PWSD | Output Shaft Power | kW | 221.8215 | 221.8214 | $6.8789\times10^{-7}$ |
| SFC | Specific Fuel Consumption | kg/(kW.h) | 0.3059 | 0.3059 | $4.8712\times10^{-7}$ |
| SNOx | NOx Severity Factor | / | 0.1134 | 0.1134 | $3.8105\times10^{-6}$ |
| HPCSM | Compressor Stability Margin | / | 29.5449 | 29.5450 | $3.0988\times10^{-6}$ |
| T1 | Inlet Total Temperature | K | 264.8290 | 264.8290 | $1.5025\times10^{-15}$ |
| P1 | Inlet Total Pressure | kPa | 35.5998 | 35.5998 | $3.9918\times10^{-16}$ |
| T2 | Compressor Inlet Total Temperature | K | 264.8290 | 264.8290 | $1.5025\times10^{-15}$ |
| P2 | Compressor Inlet Total Pressure | kPa | 48.8945 | 48.8945 | $4.3597\times10^{-16}$ |
| W2 | Compressor Inlet Flow Rate | kg/s | 1.6817 | 1.6817 | $1.3468\times10^{-6}$ |
| T3 | Compressor Outlet Total Temperature | K | 534.2960 | 534.2970 | $1.8278\times10^{-6}$ |
| P3 | Compressor Outlet Total Pressure | kPa | 408.2189 | 408.2197 | $1.9437\times10^{-6}$ |
| Ps3 | Compressor Outlet Static Pressure | kPa | 390.1103 | 390.1111 | $1.9557\times10^{-6}$ |
| W3 | Compressor Outlet Flow Rate | kg/s | 1.6817 | 1.6817 | $1.3468\times10^{-6}$ |
| T4 | Combustion Chamber Outlet Total Temperature | K | 1071.1870 | 1071.1880 | $9.1166\times10^{-7}$ |
| P4 | Combustion Chamber Outlet Total Pressure | kPa | 395.4857 | 395.4865 | $1.8520\times10^{-6}$ |
| W4 | Combustion Chamber Outlet Flow Rate | kg/s | 1.5181 | 1.5181 | $1.3349\times10^{-6}$ |
| T41 | Turbine Inlet Total Temperature | K | 1044.8800 | 1044.8800 | $2.3937\times10^{-15}$ |
| W41 | Turbine Inlet Flow Rate | kg/s | 1.6022 | 1.6022 | $1.2649\times10^{-6}$ |
| T5 | Turbine Outlet Total Temperature | K | 645.7930 | 645.7930 | $3.5208\times10^{-16}$ |
| P5 | Turbine Outlet Total Pressure | kPa | 39.1223 | 39.1223 | $7.8006\times10^{-7}$ |
| W5 | Turbine Outlet Flow Rate | kg/s | 1.6863 | 1.6863 | $1.3432\times10^{-6}$ |
| T8 | Engine Outlet Total Temperature | K | 645.7930 | 645.7930 | $3.5208\times10^{-16}$ |
| P8 | Engine Outlet Total Pressure | kPa | 37.7113 | 37.7112 | $8.0924\times10^{-7}$ |
| W8 | Engine Outlet Flow Rate | kg/s | 1.6863 | 1.6863 | $1.3432\times10^{-6}$ |
| Average Relative Error：$1.1021\times10^{-6}$ | | | | | |

6.1.2. Transients and Fault Simulation Verification for Gas Generator

Next, the transients and fault states of the gas generator model module are simulated and verified. First, the following transient process is selected and simulated using GasTurb software: the initial working state of the gas generator is ground level, 230kW shaft power, 1% compressor efficiency degradation, 2% turbine efficiency degradation, 3% compressor flow degradation, 4% turbine flow degradation, with other parameters the same as the design point. At the 3s, the gas generator input fuel flow steps up by 10%, with other parameters remaining unchanged, and the simulation runs until the 10s. Throughout the simulation process, the gas generator is connected to GasTurb's default load, where the load power is proportional to the cube of the speed. Afterward, the constructed gas generator module is used to simulate the same transient process described above, using the same load as GasTurb.

The simulation results show that the average relative error between GasTurb and the constructed gas generator module is $8.4317\times10^{-5}$. The comparison of key parameters between the two gas generator models is shown in Figure 15 to Figure 17 below. From the parameter variation comparison charts and the average relative error, it can be seen that the simulation calculation results of the constructed gas generator module in transients and fault states are almost identical to GasTurb, thus verifying the effectiveness of the gas generator module.

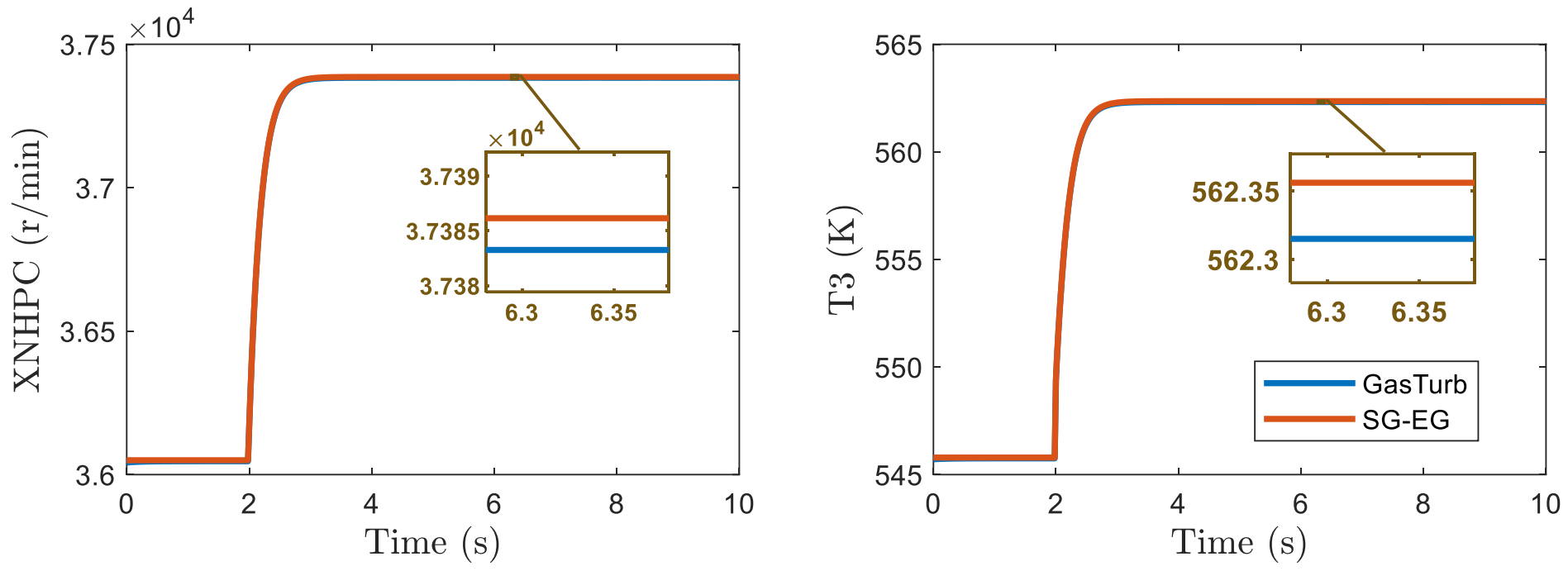

**Figure 15.** Comparison of shaft speed and compressor outlet total temperature transients between gas generator model and GasTurb model under gas path fault

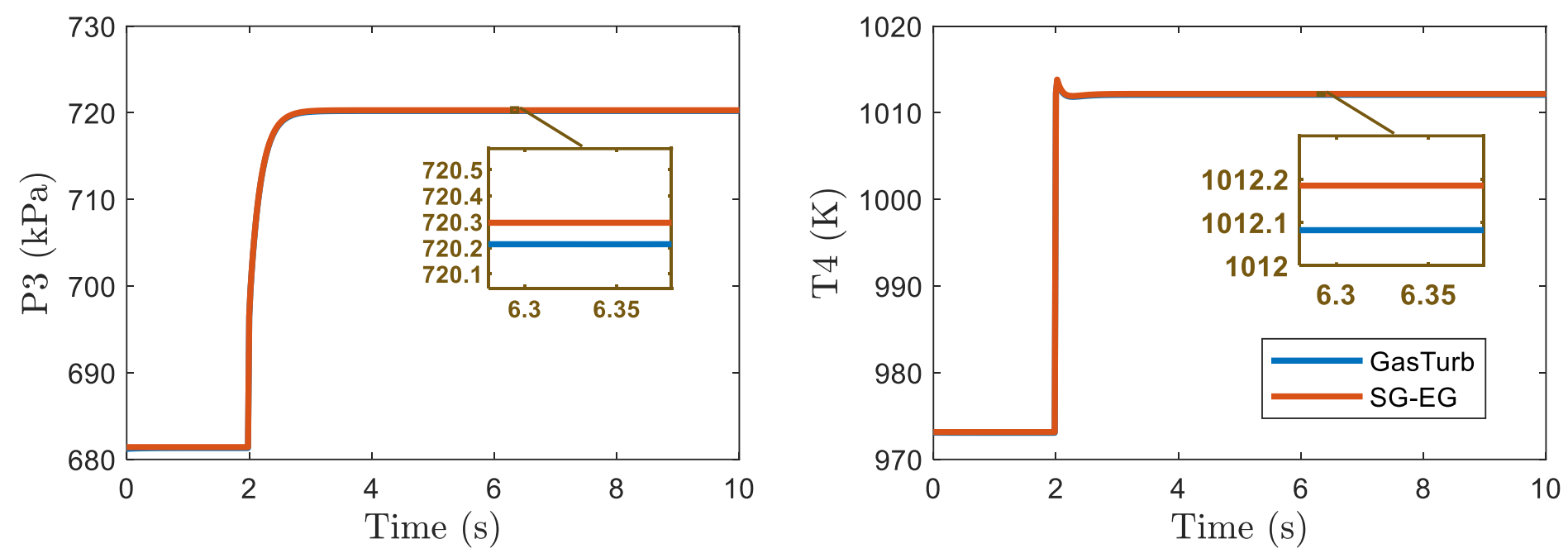

**Figure 16.** Comparison of compressor outlet total pressure and combustion chamber outlet total temperature transients between gas generator model and GasTurb model under gas path fault

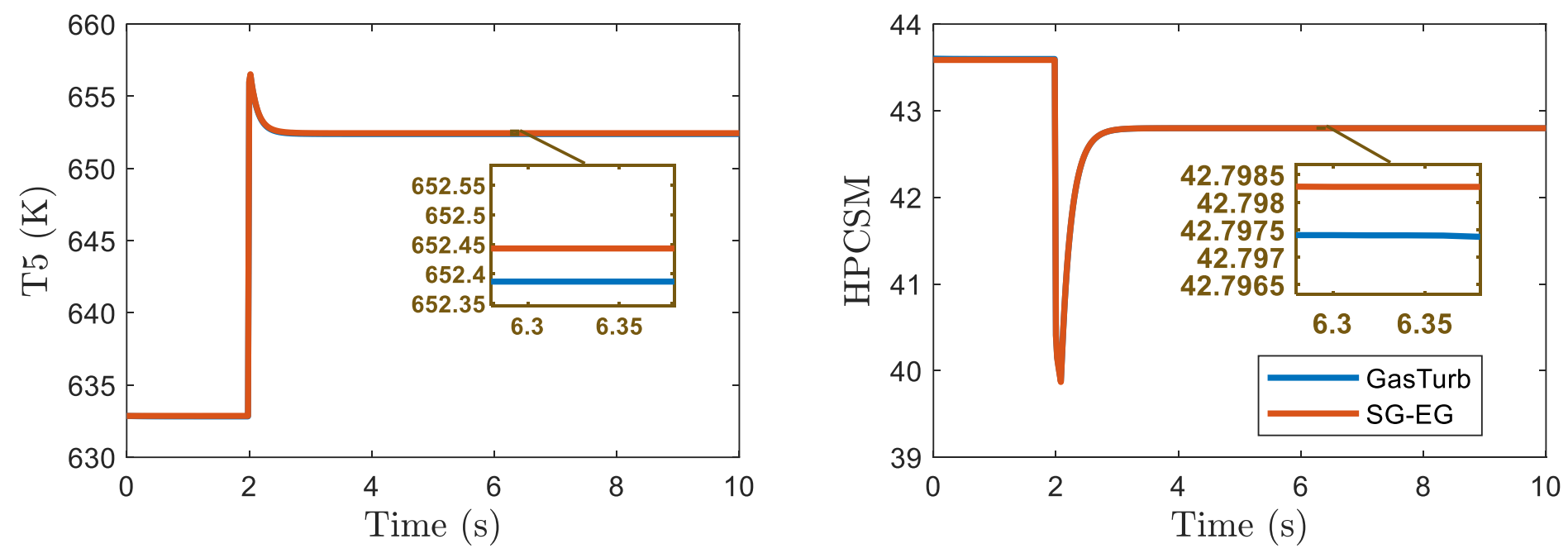

**Figure 17.** Comparison of turbine outlet total temperature and compressor surge margin transients between gas generator model and GasTurb model under gas path fault

The relative errors between the constructed gas generator model and the GasTurb software calculation results at design points, off-design points, transients and gas path fault simulations are summarized in the following Table 6. These studies demonstrate the effectiveness of the gas generator model, which can effectively simulate and model the healthy and gas path fault working conditions of the gas generator, providing support for subsequent research.

**Table 6.** Relative error table between gas generator model and GasTurb model

| Design Point | Off-Design Point | Transients and Gas Path Faults |
| --- | --- | --- |
| $1.0285\times10^{-6}$ | $2.3461\times10^{-5}$ | $8.4317\times10^{-5}$ |

*6.2. Starter/Generator Model Simulation and Verification*

In this part, first, for the starter/generator model, design point single-point simulation and off-design point multi-point simulation will be conducted, and the results will be verified using MATLAB's synchronous motor module; next, transient simulation of the starter/generator module in the healthy state will be conducted, also using MATLAB's synchronous motor module for verification; finally, transient simulation of the starter/generator in the fault state will be conducted, and the results will be verified using data from relevant literature.

#### 6.2.1. Design Point and Off-Design Point Simulation and Verification for Starter/Generator

First, for the design point of the starter/generator module, 225kW output electrical power, simulation is conducted, and at the same time, using MATLAB's synchronous generator module, a synchronous motor model with the same design parameters and design point operating conditions is established and simulated. The simulation results of both, comparison, and errors are shown in Table 7.

From Table 7, it can be seen that the design point simulation calculation results of the constructed gas generator module are almost identical to the calculation results of MATLAB synchronous generator module, with an average relative error of $6.2123\times10^{-5}$, a maximum relative error of $1.7722\times10^{-4}$ for phase A phase voltage, and a minimum relative error of $5.9752\times10^{-6}$ for phase B current. Therefore, this verifies the design point effectiveness of the starter/generator module.

**Table 7.** Design point: 225kW

| Parameter | Unit | SM MATLAB | SG Module | Relative Error |
|---|---|---|---|---|
| Phase A Voltage | V | 229.6704 | 229.7375 | $2.9225\times10^{-4}$ |
| Phase B Voltage | V | 229.6903 | 229.6715 | $8.1933\times10^{-5}$ |
| Phase C Voltage | V | 229.6704 | 229.6872 | $7.3201\times10^{-5}$ |
| AB Line Voltage | V | 397.8644 | 397.8698 | $1.3493\times10^{-5}$ |
| BC Line Voltage | V | 397.7861 | 397.7827 | $8.7435\times10^{-6}$ |
| CA Line Voltage | V | 397.8989 | 397.8970 | $4.7427\times10^{-6}$ |
| Phase A Current | A | 325.7119 | 325.7148 | $8.7442\times10^{-6}$ |
| Phase B Current | A | 325.6196 | 325.6212 | $4.7522\times10^{-6}$ |
| Phase C Current | A | 325.6478 | 325.6435 | $1.3489\times10^{-5}$ |
| Average Relative Error：$5.5705\times10^{-5}$ | | | | |

Next, for the off-design points of the gas generator module, through the way of shaft power, 10 different operating conditions were selected for simulation, and at the same time, using MATLAB synchronous generator module, simulation was conducted for engine models with the same design parameters and conditions. Due to space limitations, only the off-design point simulation results for a 100kW shaft power condition are shown, as in Table 8.

From Table 8, it can be seen that the calculation results of the constructed gas generator model and GasTurb have very small errors, with an average relative error of only $1.125\times10^{-4}$. And under all 10 selected off-design point conditions, the average error between the two parameters is $1.251\times10^{-4}$, slightly higher than the design point, but still very small. Therefore, this verifies the off-design point effectiveness of the gas generator module.

**Table 8.** Off-design point: 100kW

| Parameter | Unit | SM MATLAB | SG Module | Relative Error |
|---|---|---|---|---|
| Phase A Voltage | V | 229.9349 | 229.9623 | $1.1894\times10^{-4}$ |
| Phase B Voltage | V | 229.9349 | 229.8961 | $1.6870\times10^{-4}$ |
| Phase C Voltage | V | 229.9349 | 229.9463 | $4.9759\times10^{-5}$ |
| AB Line Voltage | V | 398.2589 | 398.2391 | $4.9748\times10^{-5}$ |
| BC Line Voltage | V | 398.2589 | 398.2116 | $1.1893\times10^{-4}$ |
| CA Line Voltage | V | 398.2590 | 398.3261 | $1.6865\times10^{-4}$ |
| Phase A Current | A | 144.8865 | 144.9038 | $1.1891\times10^{-4}$ |
| Phase B Current | A | 144.8865 | 144.8621 | $1.6868\times10^{-4}$ |
| Phase C Current | A | 144.8865 | 144.8937 | $4.9742\times10^{-5}$ |
| Average Relative Error： $1.1245\times10^{-4}$ | | | | |

6.2.2. Transients and Fault Simulation Verification for Starter/Generator

Next, the transients and fault conditions of the starter/generator module are simulated and verified. First, for the healthy condition of the starter/generator, selecting the design point condition, the three-phase current, phase voltage, and line voltage transients of the constructed starter/generator module are compared with the output of MATLAB's synchronous motor module.

The simulation results show that the average relative error of parameters between MATLAB synchronous motor module and the constructed starter/generator module is $8.4317\times10^{-5}$. The comparison of key parameter transients between the two is shown in Figure 18.

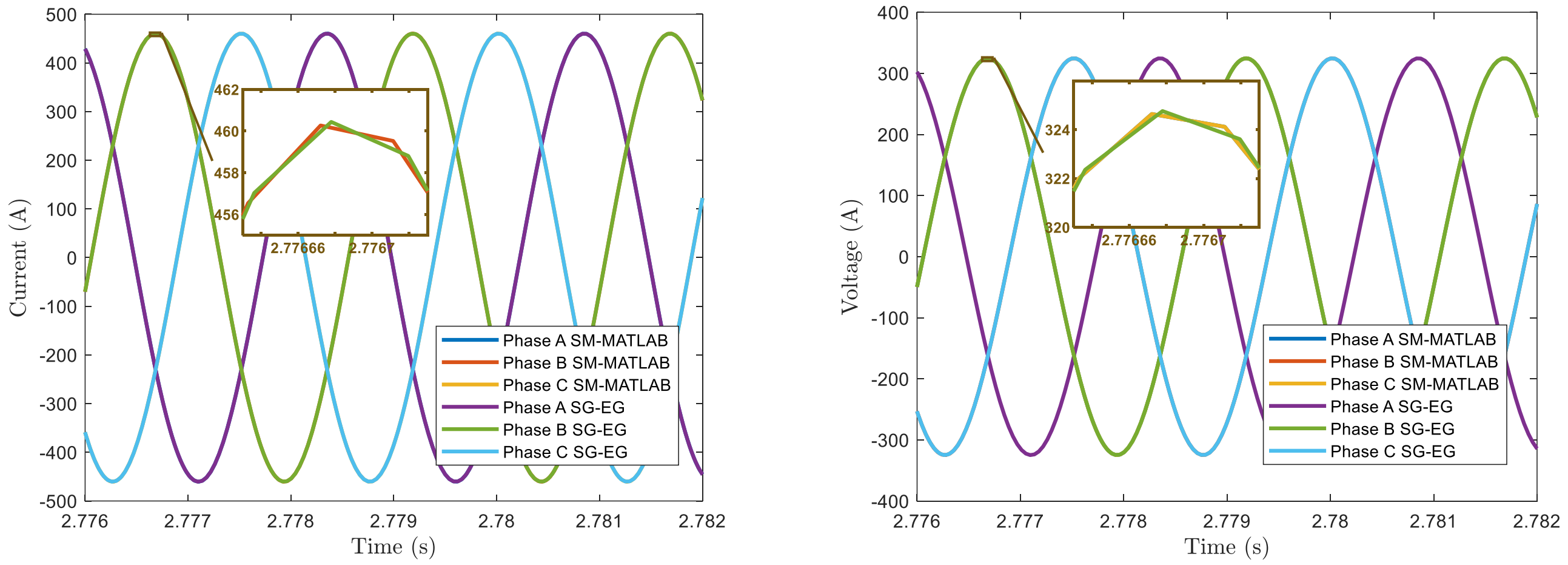

(**a**) (**b**)

**Figure 18.** Comparison of ABC three-phase current (a) and voltage (b) between starter/generator model and SM-MATLAB model

From the parameter variation comparison and the average relative error, it can be seen that the simulation calculation results of the constructed starter/generator module in the healthy condition transients are almost identical to MATLAB synchronous motor module, thus verifying the effectiveness of the starter/generator module in the healthy state.

Next, while maintaining the design point condition, a 5% stator TTSC fault is introduced into the established starter/generator module for fault transient simulation. The current changes of the starter/generator after the fault occurs are shown in Figure 19 (a). It can be observed that due to the occurrence of TTSC, changes in each phase voltage lead to the phase currents no longer maintaining a balanced state.

Reference [72] provides experimental measurement results of fault currents under 5% stator TTSC fault conditions for synchronous generators, as shown in Figure 19 (b). By comparing the experimental measurement results in reference [72] with the simulation results of the established generator model, although the numerical results are not completely consistent due to differences in generator parameters, it can be clearly seen that the fault current variation trend and patterns of the established starter/generator module is highly consistent with the results in reference [72]. This result verifies the effectiveness of the developed starter/generator module in simulating TTSC fault transient behavior.

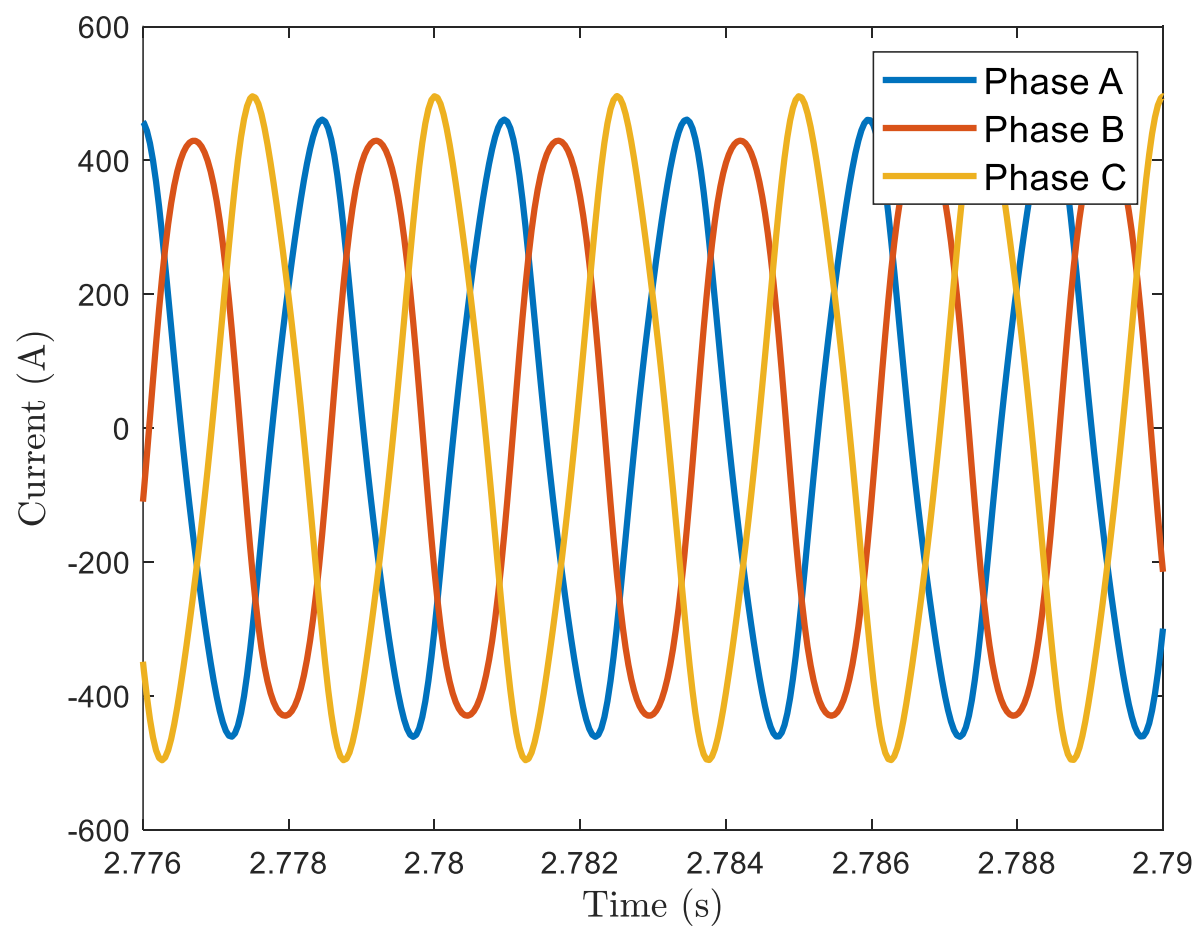

(a)

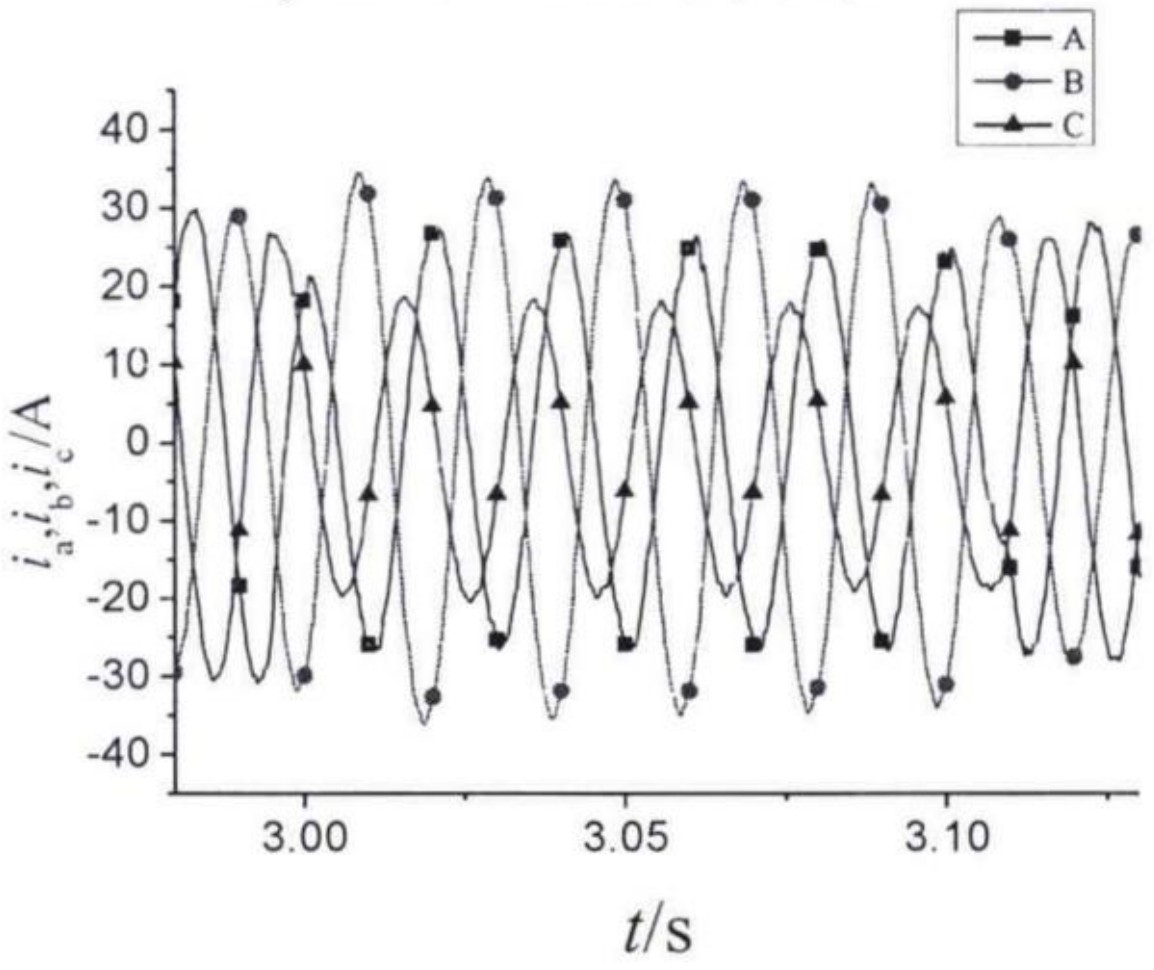

(b)

**Figure 19.** Three-phase current under TTSC fault: Simulation results from the established starter/generator model (a) compared with Experimental measurements from reference literature [72] (b)

The comparison of the constructed starter/generator model in design point, off-design point, transients and gas path fault simulations with MATLAB synchronous motor module simulation results and related literature measurement results is summarized in the following Table 9. These studies demonstrate the effectiveness of the starter/generator model, which can effectively simulate and model the healthy and TTSC fault working conditions of the starter/generator, providing support for subsequent research.

**Table 9.** Relative error table of starter/generator model

| Design Point | Off-Design Point | Healthy Transients | TTSC Transients |
|---|---|---|---|
| $7.3673\times10^{-5}$ | $1.8746\times10^{-4}$ | $5.2122\times10^{-4}$ | Consistent Patterns |

*6.3. Joint Model Simulation Verification*

The above research has verified the effectiveness of the joint model for all-electric APU starter/generator and gas generator by module. Next, the effectiveness of the entire joint simulation model will be verified.

To verify the model, a transient process involving the joint model's design point, off-design point, gas generator gas path fault, and starter/generator gas path fault was first designed, as shown in Figure 20 below.

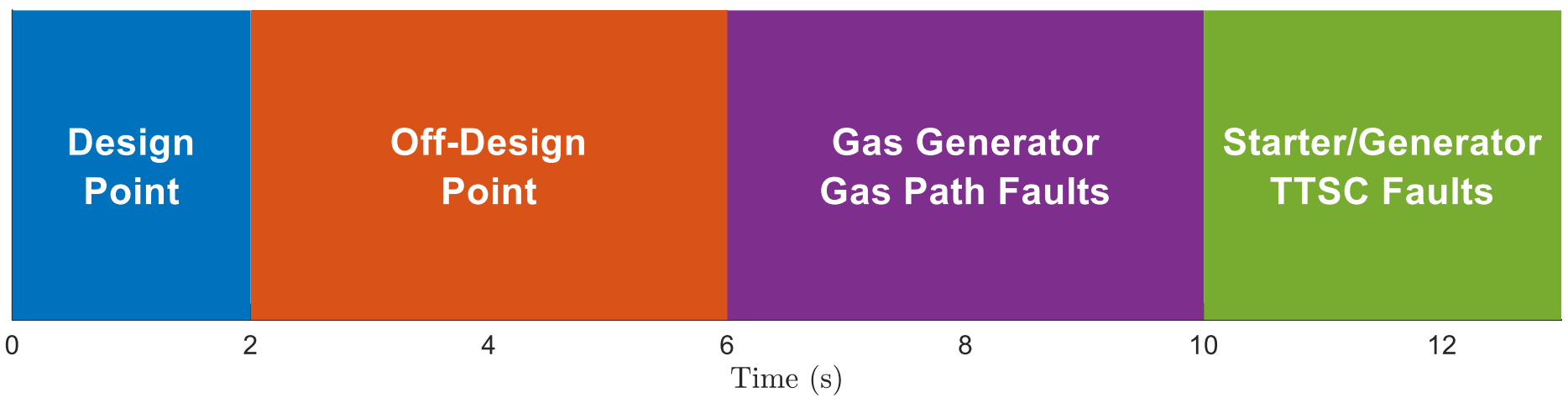

**Figure 20.** Joint model simulation phase diagram

The entire simulation process starts from the ground state, design point condition; then by unloading part of the electrical load, the system power output is reduced, conducting off-design point simulation; then gas path fault is injected into the gas generator, conducting gas generator gas path fault simulation, with fault parameters chosen the same as those in the gas generator model verification: 1% compressor efficiency degradation, 2% turbine efficiency degradation, 3% compressor flow degradation, 4% turbine flow degradation; finally, a 5% TTSC fault, the same as in the starter/generator model verification, is injected into the starter/generator, conducting starter/generator TTSC fault simulation until the end of the simulation. During the simulation process, the changes in some key parameters of the joint simulation model are shown in Figure 21 to Figure 25, including key parameter changes in the gas generator part and key parameter changes in the starter/generator part.

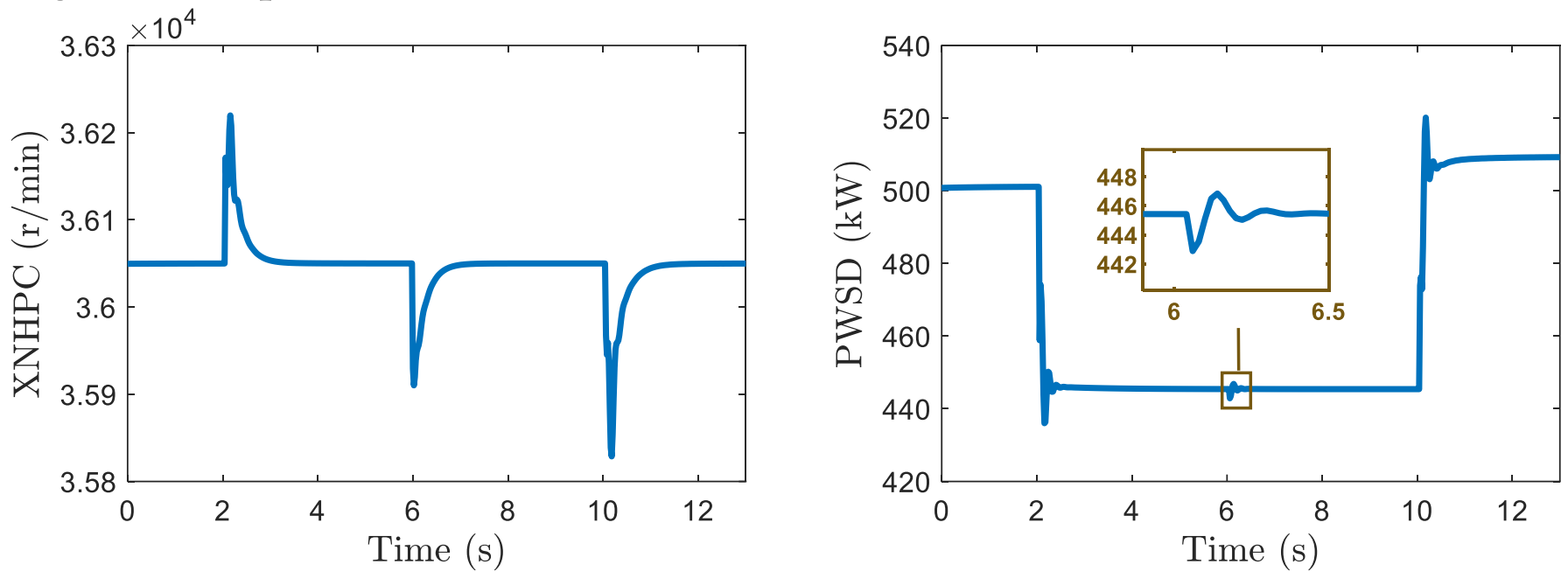

**Figure 21.** Shaft speed and shaft power transients for the gas generator during joint model simulation

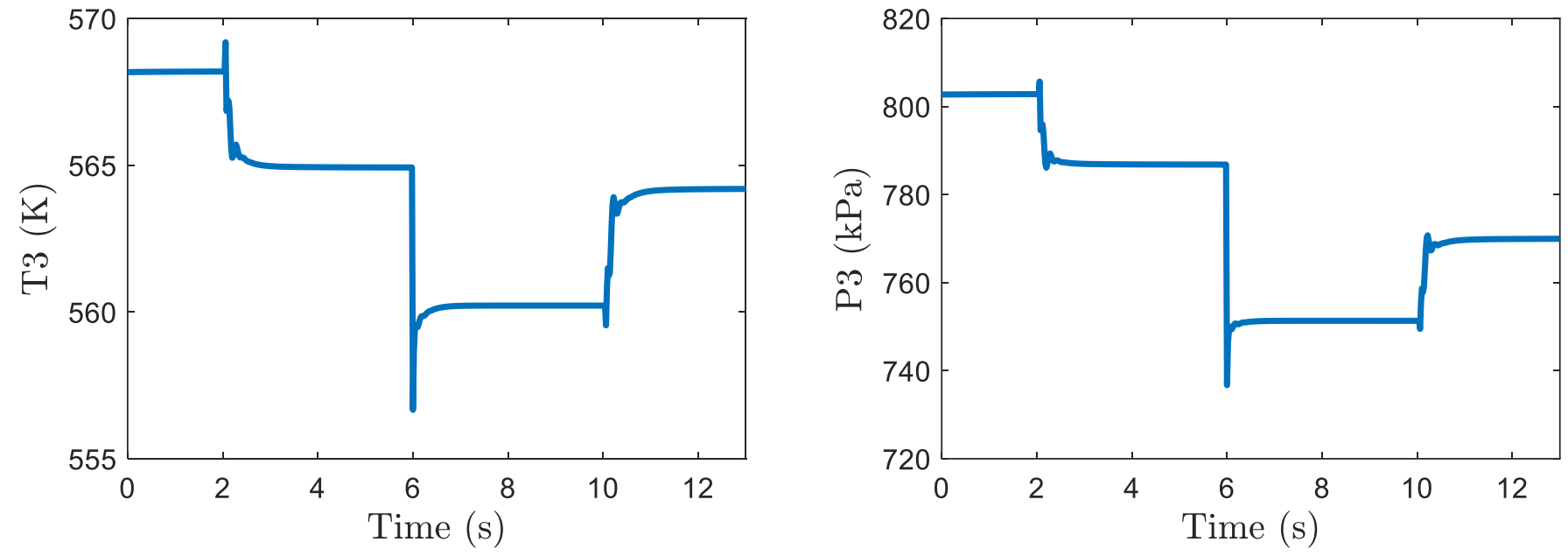

**Figure 22.** Compressor outlet total temperature and total pressure transients for the gas generator during joint model simulation

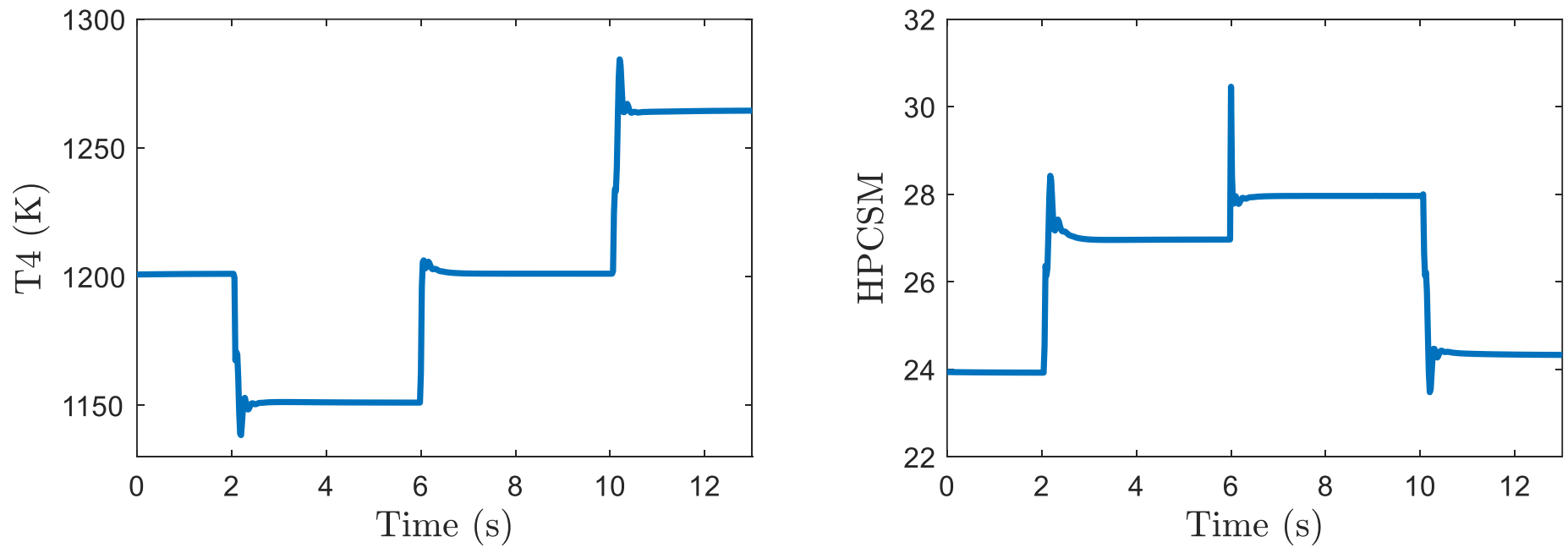

**Figure 23.** Combustion chamber outlet total temperature and compressor surge margin transients for the gas generator during joint model simulation

From the Figure 21 to Figure 25, it can be observed that when the simulation reaches the 2s, part of the electrical power is unloaded, entering off-design point simulation, the starter/generator output terminal current decreases significantly, and the terminal voltage increases, but under the control of the voltage controller regulator, it quickly recovers to the design value. At the same time, the starter/generator power decreases synchronously and rapidly, causing the gas generator output shaft power to decrease rapidly, resulting in the gas generator speed increasing rapidly. Under the adjustment of the gas generator's speed controller, the fuel flow decreases, causing the T4 temperature to decrease rapidly, the turbine power decreases, making the speed gradually recover to the design value. The entire process is also accompanied by a decrease in the compressor outlet total pressure and total temperature, and an increase in the compressor surge margin.

When the simulation reaches the 6s, the gas generator experiences gas path degradation, which will cause the compressor and turbine efficiency to decrease, the compressor flow capacity to decrease, and the turbine flow capacity to increase. Under the influence of these factors, the gas generator's ability to output power and its efficiency decrease, making the current speed no longer able to meet the required shaft power output, therefore the speed decreases rapidly. Under the adjustment of the gas generator's speed controller, the fuel flow rapidly increases, causing the T4 temperature to increase rapidly, the turbine power increases, thereby allowing the speed to gradually recover to the design value. At the same time, the speed fluctuations in this process will cause the starter/generator's speed to fluctuate, bringing about fluctuations in the output terminal voltage, but due to the action of the voltage regulator, the voltage fluctuations are not large, but still cause small fluctuations in output current, electrical power, and shaft power, which are reflected in the shaft power fluctuation enlargement in Figure 21 and the voltage and current enlargement in Figure 24 and Figure 25. The entire process is also accompanied by an increase in the compressor outlet total pressure, total temperature, and compressor surge margin.

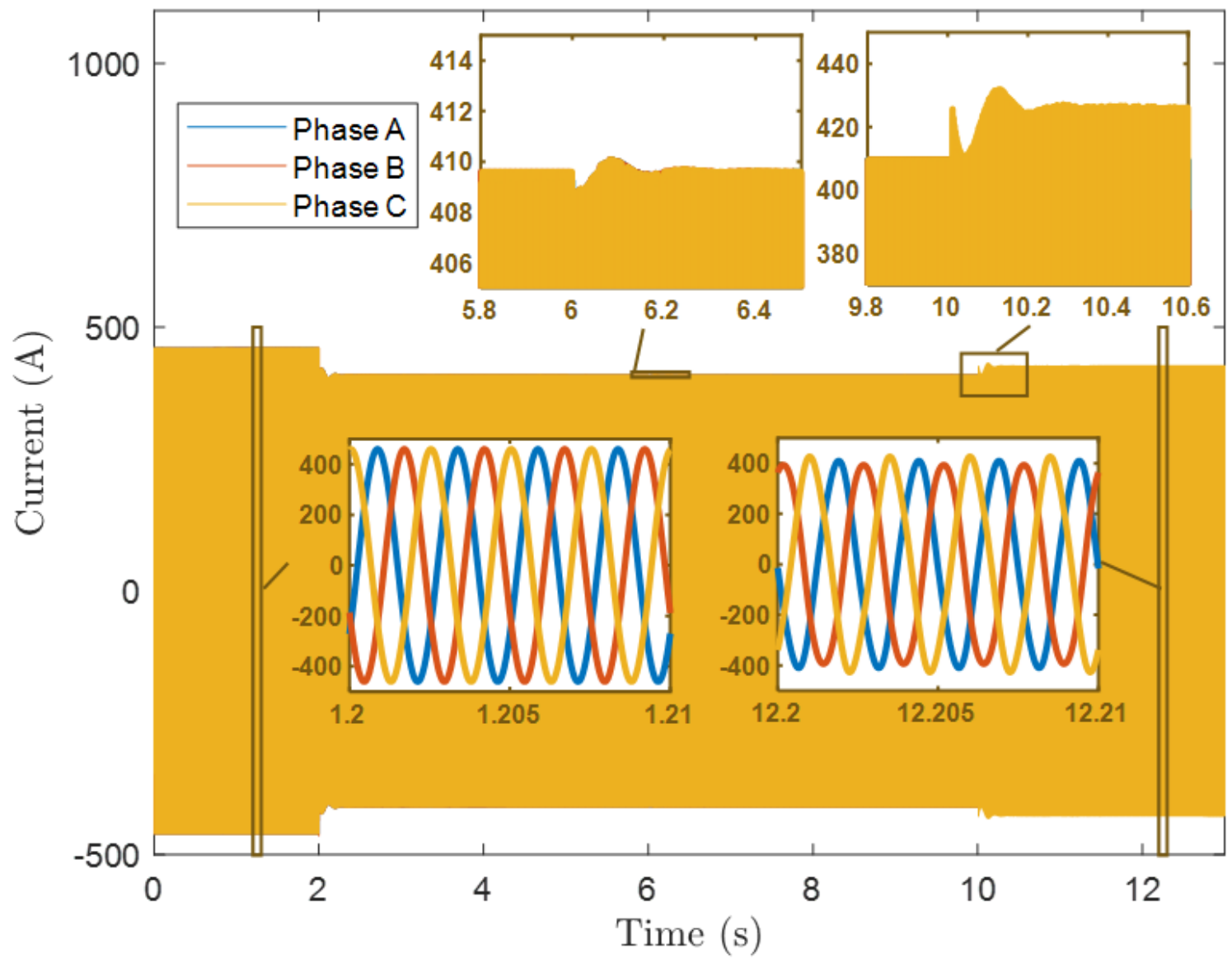

**Figure 24.** Three-phase current transients of gas generator during joint model simulation

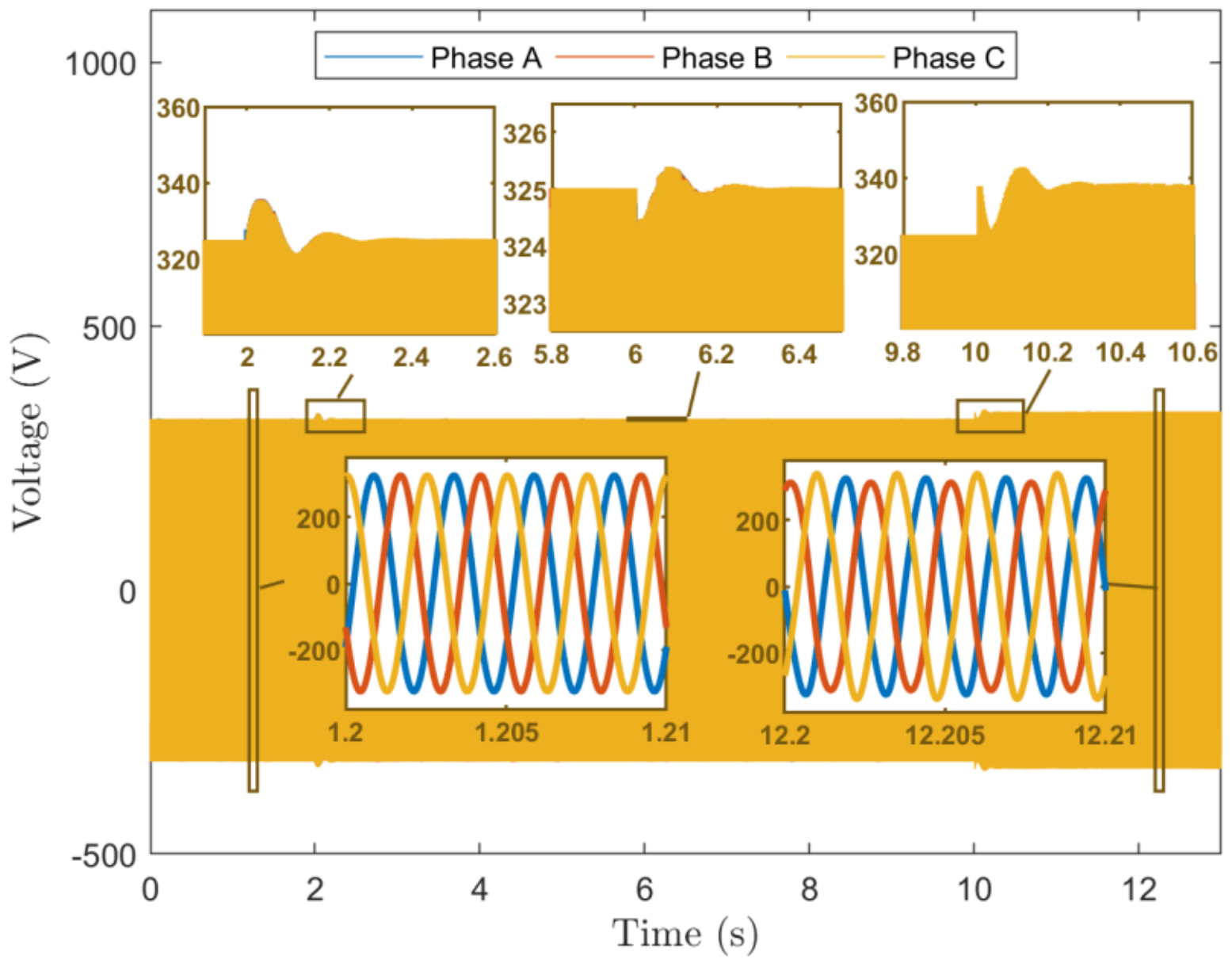

**Figure 25.** Three-phase voltage transients of gas generator during joint model simulation

When the simulation reaches the 10s, the starter/generator experiences an TTSC fault. From the current and voltage comparison in Figure 24 and Figure 25, it can be seen that the voltage and current are no longer balanced, showing high and low differences, with the fault patterns of the current consistent with the TTSC fault current patterns measured in reference [72] in Figure 19 (b). Due to the control of the voltage regulator, the voltage mean value remains near the design point, but due to the imbalance of voltage and current, some phase voltages exceed the design value, causing the voltage and current envelope to rise. Power loss and voltage changes caused by the TTSC cause the starter/generator power to increase, the gas generator shaft power to increase, further causing the gas generator speed to decrease rapidly. From Figure 24 and Figure 25, it can be seen that under the adjustment of the speed controller, fuel flow increases, T4 temperature rises, turbine power increases, making the speed gradually return to near the design value. The entire process is accompanied by an increase in the compressor outlet total temperature and total pressure, as well as a decrease in the compressor surge margin.

The parameters of the four stages in the above transient simulation process, namely the design point stage, off-design point stage, gas generator gas path fault stage, and starter/generator TTSC fault stage, are compared with the data from GasTurb gas generator model, MATLAB synchronous motor model, and related literature, with the results summarized in the following Table 10. From the table, it can be seen that the all-electric APU starter/generator-gas generator joint model constructed performs well in all stages. In cases where there is a complete reference model, the simulation results of the constructed joint model have very small errors compared to the relevant reference models, and in the TTSC stage where there is no complete reference model, the fault manifestation form of the joint model is also very close to the actual measurement data in related literature [72].

**Table 10.** Joint model fault simulation error statistics table

| Gas Generator (Compared with GasTurb) | | | Starter/Generator (Compared with SM-MATLAB and Reference [72]) | | |
|---|---|---|---|---|---|
| Design Point | Off-Design Point | Gas Path Fault | Design Point | Off-Design Point | TTSC Fault |
| $2.361\times10^{-6}$ | $2.084\times10^{-5}$ | $2.088\times10^{-5}$ | $6.628\times10^{-3}$ | $7.130\times10^{-3}$ | Consistent Patterns |

The simulation validation above extensively utilized data from GasTurb and Matlab, two well-established commercial software packages, for comparative verification. Detailed model parameters are provided to facilitate readers in reproducing and validating these simulation results. Additionally, this paper fully discloses the simulation architecture and modeling details of the joint model, enabling readers to reconstruct the proposed model, conduct systematic verification, and develop further research based on this foundation.

Through the joint model simulation and verification in this section and the module-level verification in the previous two sections, the effectiveness of the all-electric APU joint fault model constructed in this paper is fully demonstrated, showing that it can effectively simulate and model the healthy and TTSC fault working conditions of the starter/generator, providing support for subsequent research.

## 7. Conclusions

This paper presented a joint system modeling approach for fault simulation of all-electric APU starter/generators and gas generators. To address the joint fault modeling challenges of electromechanical coupling, simulation precision, and computational efficiency, we proposed three key technical innovations: First, we developed a multi-rate continuous-discrete hybrid simulation architecture that treats the starter/generator as a continuous system with variable step size in Simulink while modeling the gas generator as a discrete system with fixed step size in a DLL environment. This architecture also significantly reduced computational load without compromising accuracy. Second, we implemented a multi-loop modeling approach for the starter/generator that accurately simulates both normal operation and TTSC faults. For the gas generator, we developed an improved GasTurb-DLL modeling method (IGDM) that enhances uncertainty modeling, state-space representation, and tool chain compatibility. Validation results comparing our model with reference data from GasTurb software, MATLAB's synchronous motor module, and literature showed close agreement. The error patterns were consistent, with relative errors ranging from $2.361\times10^{-6}$ to $7.130\times10^{-3}$. This study's proposed joint fault modeling methodology establishes a basis for exploring the FDI challenges and opportunities arising from all electrification of the APU, including joint fault estimation and diagnosis, coupled electromechanical fault characteristics.

**Author Contributions:** Conceptualization, H.M.; methodology, H.M.; software, H.M.; validation, H.M.; formal analysis, H.M.; investigation, H.M.; resources, Y.G.; data curation, H.M.; writing—original draft preparation, H.M.; writing—review and editing, Y.G.; visualization, H.M.; supervision, Y.G.; project administration, Y.G.; funding acquisition, Y.G. All authors have read and agreed to the published version of the manuscript.

**Funding:** This research was funded by China Scholarship Council under Grant 201906290241.

**Data Availability Statement:** The original contributions presented in the study are included in the article, further inquiries can be directed to the corresponding author.

**Conflicts of Interest:** The authors declare no conflicts of interest.

## Abbreviations

The following abbreviations are used in this manuscript:

| | |
|---|---|
| APU | Auxiliary power unit |
| TTSC | Turn-to-turn/inter-turn short-circuit |
| IGDM | Improved GasTurb-DLL modeling method |
| FDI | Fault detection and isolation |
| MEA | More electric aircraft |
| AEA | All-electric aircraft |
| GDM | GasTurb-DLL modeling method |
| CSD | Constant speed drives |
| WRSG | Wound rotor starter/generator |
| WFA | Winding function approach |
| DLL | Dynamic-link library |
| BP | Back propagation |
| RBF | Radial basis function |